%% file: main.tex
\documentclass[twocolumn]{aastex62} 
\usepackage{amsmath,amstext}
\usepackage[figure,figure*]{hypcap}
\usepackage{subfigure,graphicx}
\usepackage{comment}
\usepackage{lmodern}
\usepackage[T1]{fontenc}

\newcommand{\msun}{\ifmmode {M_{\odot}}\else${M_{\odot}}$\fi}
\newcommand{\rsun}{\ifmmode {R_{\odot}}\else${R_{\odot}}$\fi}
\newcommand{\Msun}{\ifmmode {M_{\odot}}\else${M_{\odot}}$\fi}
\newcommand{\Rsun}{\ifmmode {R_{\odot}}\else${R_{\odot}}$\fi}
\newcommand{\Lsun}{\ifmmode {L_{\odot}}\else${L_{\odot}}$\fi}
\newcommand{\lapprox }{{\lower0.8ex\hbox{$\buildrel <\over\sim$}}}
\newcommand{\gapprox }{{\lower0.8ex\hbox{$\buildrel >\over\sim$}}}
\def\amin{\ifmmode^{\prime}\else$^{\prime}$\fi}
\def\asec{\ifmmode^{\prime\prime}\else$^{\prime\prime}$\fi}
\newcommand{\degree}{\ifmmode {^\circ}\else$^\circ$\fi}

\newcommand{\Ro}{\ifmmode {R_o}\else$R_o\ $\fi}
\newcommand{\lha}{\ifmmode {L_{H\alpha}/L_{bol}}\else$L_{H\alpha}/L_{bol}$\ \fi}

\newcommand{\tess}{TESS}
\newcommand{\ktwo}{K2}

\newcommand{\tausq}{\ifmmode {\tau^2}\else$\tau^2$\fi}

\newcommand{\prot}{\ensuremath{P_{rot}}} 

\newcommand{\agerange}{25--55~Myr}

\newcommand{\totaltargets}{1410}

\newcommand{\tessperiods}{868}

\newcommand{\tesslit}{59}
\newcommand{\totalit}{114}

\newcommand{\lithigh}{11} 
\newcommand{\duplit}{ten}
\newcommand{\newperiods}{744}
\newcommand{\newsolar}{96}

\shorttitle{{\it TESS} Periods for ZAMS stars}
\shortauthors{Douglas et al.}

\begin{document}

\title{Constraining Stellar Rotation at the Zero-Age Main Sequence with TESS}

\correspondingauthor{S.~T.~Douglas}
\email{douglste@lafayette.edu}

\author[0000-0001-7371-2832]{S. T. Douglas}
\affiliation{Lafayette College, 730 High St, Easton, PA 18042, USA}

\author[0000-0002-1617-8917]{P. A. Cargile}
\affiliation{Center for Astrophysics $\vert$ Harvard\ \&\ Smithsonian, 60 Garden St, Cambridge, MA 02138, USA}

\author[0000-0001-9590-2274]{S. P. Matt}
\affiliation{Homer L. Dodge Department of Physics and Astronomy, University of Oklahoma, Norman, OK 73019, USA}

\author[0000-0002-8228-3694]{A. A. Breimann}
\affiliation{University of Exeter, Devon, Exeter, EX4 4QL, UK}

\author[0000-0002-7407-5297]{J. A. P\'erez Ch\'avez}
\affiliation{Interdisciplinary Studies Department, Howard University,  Washington, DC 20059, USA
} 

\author[0000-0003-0918-7484]{C. X. Huang}
\affiliation{University of Southern Queensland, Toowoomba, Queensland, Australia}

\author[0000-0002-8389-8711]{N. J. Wright}
\affiliation{Astrophysics Group, Keele University, Keele, ST5 5BG, UK}

\author[0000-0002-4891-3517]{G. Zhou}
\affiliation{University of Southern Queensland, Toowoomba, Queensland, Australia}

\begin{abstract}
The zero-age main sequence (ZAMS) is a critical phase for stellar angular momentum evolution, as stars transition from contraction-dominated spin-up to magnetic wind-dominated spin-down. 
We present the first robust observational constraints on rotation for FGK stars at $\approx40$~Myr.
We have analyzed TESS light curves for \totaltargets\ members of five young open clusters with ages between 25-55~Myr: IC~2391, IC~2602, NGC 2451A, NGC 2547, and Collinder~135. 
In total, we measure \tessperiods\ rotation periods, including \newsolar\ new, high-quality periods for stars around 1~\Msun. This is an increase of \duplit\ times the existing literature sample at the ZAMS. 
We then use the \tausq\ method to compare our data to models for stellar angular momentum evolution. 
Although the ages derived from these rotation models do not match isochronal ages, we show these observations can clearly discriminate between different models for stellar wind torques.
Finally, \tausq\ fits indicate that magnetic braking and/or internal angular momentum transport significantly impact rotational evolution even on the pre-main sequence. 
\end{abstract}

\keywords{open clusters: individual (IC 2391, IC 2602, NGC 2451A, NGC 2547, Collinder 135) ---
stars:~evolution -- stars:~late-type -- stars:~rotation}

\section{Introduction}\label{intro}

The Zero-Age Main Sequence (ZAMS) is a critical point in the evolution of stellar rotation.
Low-mass stars are born with a range of rotation rates, and as they contract on the PMS, they spin up to conserve angular momentum \citep[e.g.,][]{barnes2003,bouvier2014}.
Two processes --- contraction and braking due to stellar winds --- dominate the evolution of stellar rotation between $\approx$10--100~Myr, making a significant impact on the \prot\ distributions of stars at the ZAMS.
Contraction dominates a star's rotational evolution on the pre-main sequence (PMS); stellar wind torques cannot shed enough angular momentum to prevent this spin-up. 
At the ZAMS, contraction ends and braking by the magnetized stellar wind begins to spin the star down.
This critical transition point, however, is largely unconstrained for solar-type stars with masses between $0.9-1.1$~\Msun.

The precise shape of angular momentum evolution tracks around the ZAMS is affected by angular momentum transport between the radiative core and convective outer envelope, and stars with different initial \prot\ behave differently before and after they peak \citep{gallet2015,garraffo2018,gossage2021}.
Initially rapid rotators experience the largest relative decrease in \prot\ on the PMS, and then must slow down significantly as well once they reach the ZAMS.
Conversely, slower rotators do not spin up as much, and experience weaker braking on the main sequence.
Clusters around the ZAMS still exhibit a wide spread of \prot\ for solar-mass stars, meaning we must measure \prot\ for as many stars as possible to determine the full distribution of periods at this age.

The last decade has seen an explosion of \prot\ measurements in open clusters, but there are still very few \prot\ for solar-type stars around the ZAMS.
Instead, we rely on h Per (13~Myr) and the Pleiades (125~Myr) to constrain behavior on either side of the ZAMS.

Pre-TESS studies of rotation in nearby ZAMS clusters have been stymied by their large angular sizes and the fact that many of them lie in the Southern Hemisphere.
Only $\alpha$ Per (71--85~Myr) is completely visible in the Northern sky, but its members are also sparsely distributed. 
Recently, there have been $\approx$20 \prot\ measured in the $\beta$ Pic moving group (20~Myr) and $>$200 in $\alpha$~Per, but only a dozen \prot\ are for solar-type stars \citep[][]{messina2017-1,boyle2023}. 
Between $25-55$~Myr, there are many \prot\ measurements for K and M dwarfs---but just eleven for solar-type stars  \citep[][]{patten1996, patten1996-1, barnes1999,irwin2008,messina2011}.
These existing \prot\ measurements are insufficient to constrain the rotational evolution of solar-type stars as they reach the ZAMS and begin to spin down. 


Theoretical work has been stymied by this lack of empirical constraints, because model tracks for the \prot\ minimum near the ZAMS are complex and depend on both \prot\ and stellar mass.
For example, \citet{gallet2013,gallet2015} use the 13 Myr-old h Per cluster, along with K and M dwarf members of the $\approx$35 Myr-old NGC 2547 cluster, to constrain their solar-mass model at the ZAMS.
Given that they also find that core-envelope coupling time dramatically increases for lower-mass stars, K and M dwarfs are insufficient for constraining the evolution of more massive F and G stars.
We therefore cannot be sure that the model physics is correct without robust observational data for solar-type stars.

\begin{figure*}[t]
\centerline{\includegraphics[width=\textwidth]{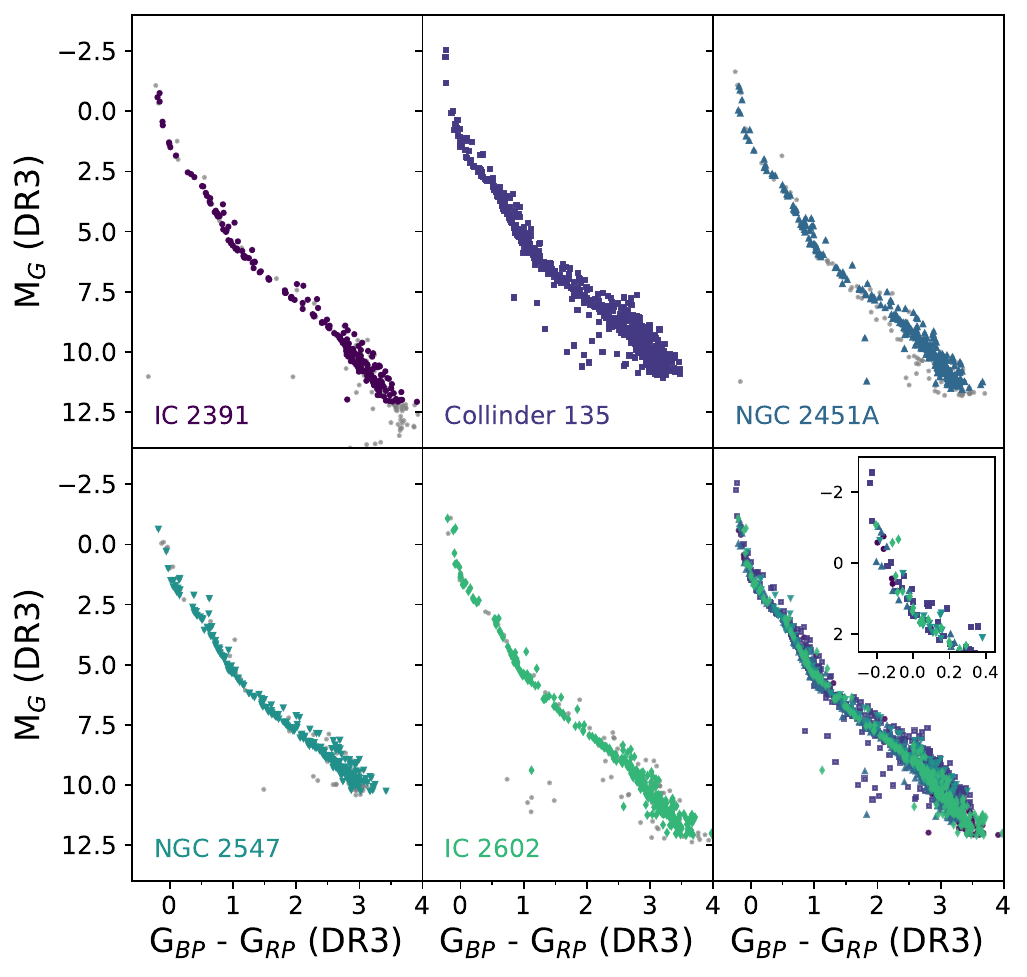}}
\caption{Gaia DR3 Color-magnitude diagram for our five target clusters. In the individual cluster panels, grey points indicate members from any of our three input catalogs, while colored points indicate members in at least two catalogs (retained for further analysis). 
Note that Collinder~135 was only included in two input catalogs, so we retain all candidate members.
The sixth panel shows the selected members of all five clusters, with an inset axis zooming in on the main sequence turnoff. 
All five clusters have indistinguishable main sequences, and lack sufficiently evolved stars that could provide stronger constraints on the cluster age. 
This supports our decision to combine the five clusters into a single population.
}
\label{fig:cmds}
\end{figure*}

\begin{deluxetable}{lrrrr}
\tabletypesize{\footnotesize}
\tablewidth{0pt}
 \tablecaption{Overlap between catalogs for each cluster
 \label{tab:memb_overlap}}
 \tablehead{
 \colhead{Catalog}  & \colhead{HDBScan} & \colhead{GES} & \colhead{CG2020} & \colhead{Unique\tablenotemark{a}} 
 }
 \startdata
 \cutinhead{Collinder 135\tablenotemark{b}}
 HDBScan & \textbf{1115} & \nodata & 314 & 801\\
 GES & \nodata & \nodata & \nodata & \nodata \\
 CG2020 & 314 & \nodata & \textbf{329} & 13 \\
 \cutinhead{IC 2391}
  HDBScan & \textbf{313} & 37 & 216 & 90 \\
 GES & 37 & \textbf{43} & 30 & 4 \\
 CG2020 & 216 & 30 & \textbf{228} & 10 \\
 \cutinhead{IC 2602}
  HDBScan & \textbf{563} & 50 & 305 & 251 \\
 GES & 50 & \textbf{54} & 43 & 4 \\
 CG2020 & 305 & 43 & \textbf{318} & 13 \\
 \cutinhead{NGC 2451A}
 HDBScan & \textbf{541} & 41 & 332 & 205\\
 GES & 41 & \textbf{42} & 37 & 1 \\
 CG2020 & 332 & 37 & \textbf{338} & 3\\
 \cutinhead{NGC 2547}
 HDBScan & \textbf{502} & 169 & 231 & 229 \\
 GES & 169 & \textbf{181} & 128 & 8 \\
 CG2020 & 231 & 128 & \textbf{232} & 0 \\
 \enddata
 \tablenotetext{a}{Members found only in the catalog for this row} 
 \tablenotetext{b}{Not included in \citet{jackson2020}} 
\end{deluxetable}

Given the dramatic improvement in astrometric cluster membership probabilities based on Gaia mission data, as well as improvements in both ground- and space-based photometry, the time is right for revisiting and expanding the \prot\ catalog at the ZAMS.
We now present \prot\ for five ZAMS clusters in the Southern sky: Collinder 135, NGC 2451A, IC 2391, IC 2602, and NGC 2547, all around \agerange\ old. 
We discuss cluster membership catalogs and existing rotation data in Section~\ref{data}, and calculate additional stellar data in Section~\ref{prop}.
We present TESS data and the resulting \prot\ values in Sections~\ref{prot} and \ref{res}. 
We fit these \prot\ to models for stellar angular momentum evolution using the \tausq\ method, a two-dimensional goodness of fit statistic previously described in \citet{naylor2006,breimann2021}. We discuss the results of these fits in Section~\ref{tausq}. 
We conclude in Section~\ref{concl}. 

\section{Existing Data}\label{data}
\subsection{Membership}\label{memb}

We combine three membership catalogs to produce a single catalog for each open cluster. We use our own membership selection as the base catalog, using the latest Gaia astrometry. For each cluster, we perform a cone-search within Gaia DR3 \citep{gaiadr3} centered on the sky coordinates for each cluster and using a search radius defined by the cluster properties given in \citet{cantat2020}. Specifically, we use a cone-search radius of three times the radius defined by \citeauthor{cantat2020} as containing half the members. We use the hierarchical clustering algorithm HDBSCAN \citep{McInnes2017} to identify over-densities in the number of stars with similar proper motions and parallaxes. For each set of data from the individual cone-searches, we select the grouping of points associated with the cluster by identifying the most stable over-density. 
All stars that are part of that over-density are considered cluster members.
We note that our membership catalog contains systematically more stars than the catalogs given in \citet{cantat2020}. This is primarily due to their use of a faintness limit of $G = 18$ mag in their selection of Gaia stars, whereas our membership criterion does not include a magnitude cut. 
 
\citet{cantat2020} identify cluster members using the unsupervised classification scheme UPMASK \citep{krone2014,cantat2018} and Gaia DR2 astrometry.
We select members from this catalog with $P_{mem}\ \geq\ 0.7$ based on those authors' calculations. 

\citet{jackson2020} combine Gaia DR2 astrometry with spectroscopy from the Gaia-ESO survey (GES), improving membership selection for stars with radial velocities. 
We select GES members with $P_{mem}\ \geq\ 0.9$, based on those authors' calculations.  

Our primary analysis only uses stars found in two or more catalogs. 
\citet{jackson2020} do not include Collinder 135 in their survey, so for that cluster we only use HDBScan and \citet{cantat2020} membership. 
The overlap between these three catalogs for each cluster are presented in Table~\ref{tab:memb_overlap}, and the final membership selections are shown in Figure~\ref{fig:cmds}. 
We measure \prot\ for all members from all three catalogs (see Section~\ref{prot}), and include membership flags in the final results.

\begin{figure}[t!]
\centerline{\includegraphics[width=3.5in]{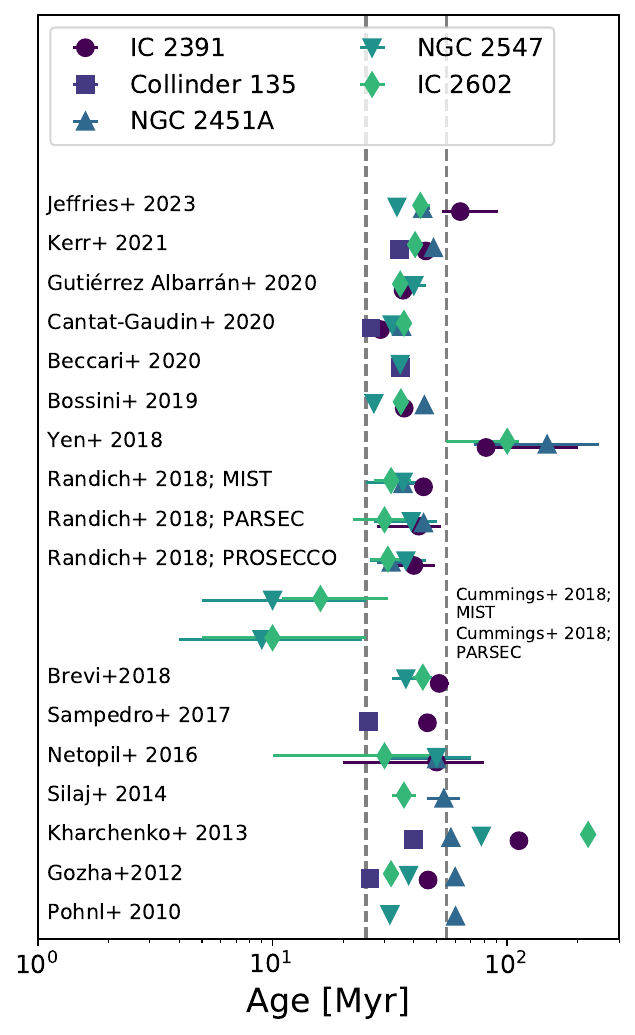}}
\caption{Ages for our five target clusters, for any papers since 2010 including two or more of our clusters. 
Older papers are at the bottom, progressing to newer results at the top.
Errorbars indicate the age uncertainty from papers that include it; many do not list an uncertainty. Some papers include multiple age determinations, which are shown separately. Particularly since Gaia data became available, the ages for all five clusters have mostly settled into the range 25--55~Myr (indicated by grey dashed lines). However, the relative ages of the five clusters are not well-determined. We therefore consider all clusters as falling into within \agerange, with no attempt to order them by age. 
}
\label{fig:ages}
\end{figure}

\subsection{Cluster ages}\label{cluster_props}

We compile measurements of cluster properties from the literature since 2010. This cut-off was chosen to eliminate older studies using less precise proper motions.
Figure~\ref{fig:ages} includes ages from papers that include at least two of our target clusters, showing the general agreement and occasional scatter among other studies. 
We summarize the literature parameters below. 

In general, most literature studies place all five of our target clusters between \agerange. 
Only one pre-Gaia study includes all five targets:
\citet{kharchenko2013}. These authors use PPMXL \citep{ppmxl} and 2MASS \citep{2mass} data to derive cluster membership and properties.\footnote{\citet{gozha2012} also includes all five, but they compile properties from the literature instead of deriving their own. Their average values are included in Figure~\ref{fig:ages}.} 
Their results for Collinder~135 (40~Myr) and NGC~2451A (58~Myr) are within the range of other literature ages for these clusters. 
Their results for IC~2602 (221~Myr), NGC~2547 (77~Myr), and IC~2391 (112~Myr), however, are completely inconsistent with any other studies of those clusters, as seen in Figure~\ref{fig:ages}.

In the Gaia era, 
\citet{cantat2020} include all five clusters and give ages between 26-36~Myr. 
\citet{bossini2019} and \citet{randich2018} include four of our five targets (neither study includes Collinder~135). 
\citet{bossini2019} gives a range of 27-44~Myr while \citet{randich2018} gives a range of 30-44~Myr for IC~2602, NGC~2547, IC~2391, and NGC~2451A. 
\citet{randich2018} derive ages using three different evolutionary models, and all three models place IC~2602, NGC~2547, and IC~2391 in the same age order. 
NGC~2451A, however, appears at a different point in the age order with each model (second-youngest at 32~Myr with PROSECCO, oldest at 44~Myr with PARSEC, and in the middle at 36~Myr with MIST). 
Furthermore, the age ordering from \citet{bossini2019} is completely different, with NGC~2547 as the youngest, IC~2602 and IC~2391 $\sim$10~Myr older, and NGC~2451A as the oldest cluster. 
As Figure~\ref{fig:ages} shows, no two studies agree on the age ordering among our target clusters, although the few studies that include Collinder~135 tend to place it at the youngest end.

Due to the inconsistency of literature results, we assume that all clusters within our sample are between \agerange, but we do not adopt individual cluster ages. 
While there are a few outliers \citep{kharchenko2013,Cummings2018,yen2018}, the majority of literature results are within this age range. 
Results from individual studies may depend on the membership catalogs used and/or the stellar evolution models chosen \citep[e.g.,][]{randich2018}. 
In addition, NGC~2547 and Collinder~135 have been linked as part of the $\approx35$~Myr-old Theia~74/Vela-CG4 filament \citep{kounkel2020,beccari2020,kerr2021}. 
We therefore combine all of our clusters into a single sample spanning a range of possible ages, rather than attempting to place them at specific ages individually.

\subsection{Existing Rotation Catalogs}\label{lit}

There have been limited photometric rotation surveys in these young ZAMS clusters, due to their southern declinations and sparse distribution on the sky. Some spectroscopic surveys have been conducted, but photometric rotation periods offer a more direct measurement with fewer inclination effects. We therefore focus our literature survey and current analysis on photometric \prot\ only. 

In this section, we review the existing literature \prot\ and match those surveys to our membership catalogs. 
We use the \texttt{simbad} and \texttt{vizier} modules in the \texttt{astroquery} package to match older catalog names to TIC and Gaia DR2 identifiers. Valid Simbad identifiers, TIC IDs, and Gaia DR2 IDs are provided in Table~\ref{tab:lit_periods} along with the original source names and periods.

\begin{figure}[!t]
\centerline{\includegraphics[width=\columnwidth]{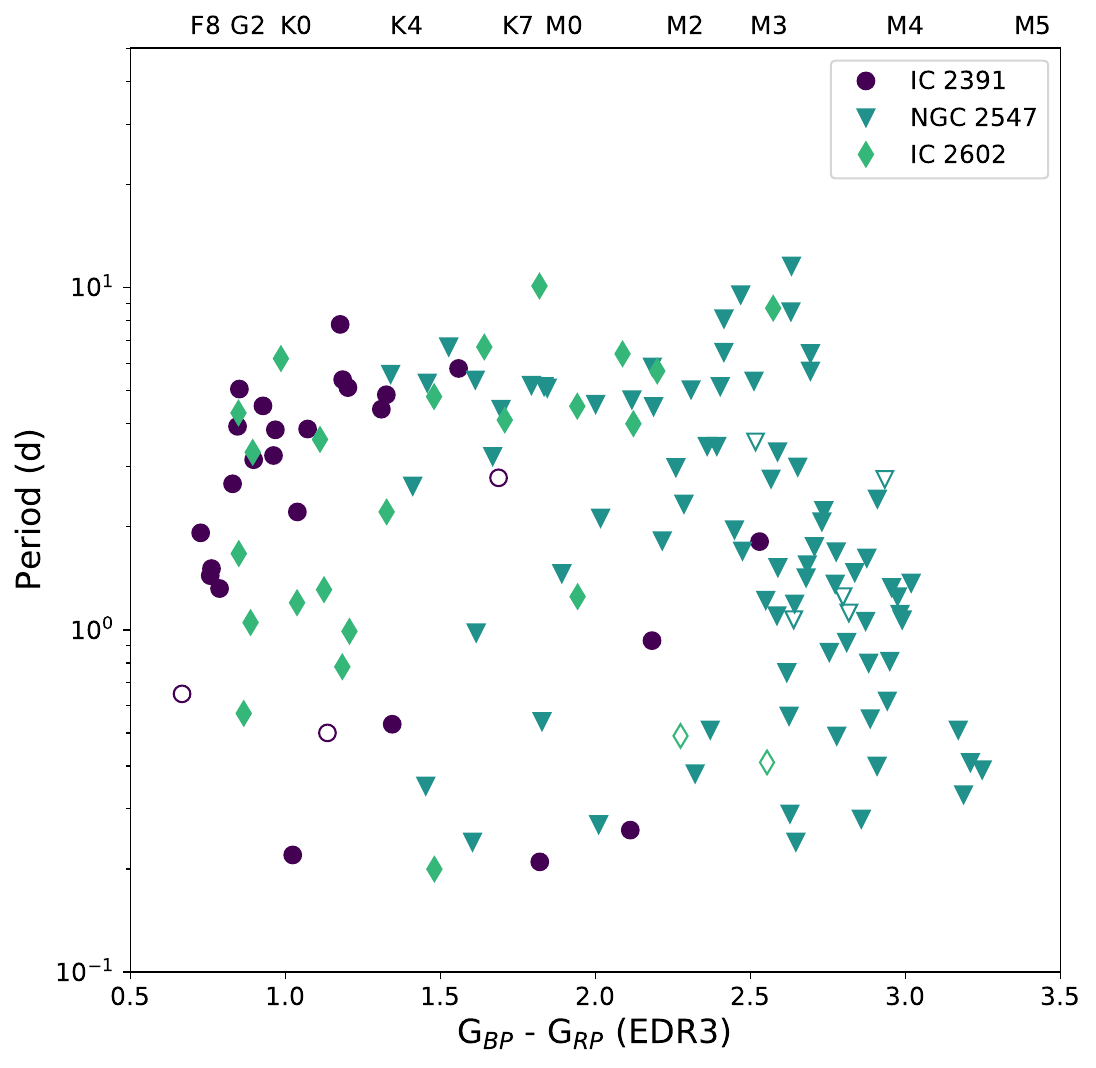}}
\caption{Literature \prot\ for our 50-Myr-old target clusters, with approximate spectral types for reference. Data for IC 2391 \citep{patten1996} are shown as purple dots, for IC 2602 \citep{barnes1999,tschape2001,patten1996-1} as blue triangles, and NGC 2547 \citep{irwin2008} as green diamonds. Open symbols indicate stars in only one of our input membership catalogs; solid symbols indicate stars in two or three input catalogs, which are used for the remaining analysis. 
There is a significant amount of data for low-mass stars in NGC 2547, but we lack strong constraints for Sun-like stars.
}
\label{fig:periodcolor_lit}
\end{figure}

\citet{patten1996} observed IC 2391 and measured \prot\ for 16 candidate cluster members with spectral types F8-M3. Their data consists of an 11-night observing run with 5 observations per night in 1993, and a 16-night run with 3-4 observations per night in 1994; only five stars were observed in both years.  All rotators in their catalog have $P_{rot}\ <\ 6$~days, and almost half have $P_{rot}\ <\ 1$~days. 
Fifteen rotators from \citet{patten1996} are also in our membership catalog for IC 2391 and have TESS data.

\input{tab_all_lit_periods}

\citet{messina2011} use photometry from the All Sky Automated Survey (ASAS) and SuperWASP survey to measure \prot\ for a variety of young stars, including 21 in IC~2391. 
Their sample spans roughly $0.5\ <\ B-V\ <\ 1.3$ (spectral types late F to mid K), and includes mostly slow rotators. 
Seven of their targets overlap with \citet{patten1996}, all with consistent \prot\ values. 
These authors provide quality flags for their \prot\ values, but we find that even their ``uncertain'' \prot\ generally match the TESS periods we derive below. 
Twenty rotators from \citet{messina2011} are also in our membership catalog for IC 2391 and have TESS data.

\citet{barnes1999} observed IC 2602 and present \prot\ for 33 stars from \citet{stauffer1997} 
with $0.5\ <\ B-V\ <\ 1.6$ (spectral types late F to mid M). 
The stars were observed over three non-contiguous observing runs in 1995: roughly once per night for a week, 3-4 times per night for three weeks, and an unstated cadence for another two weeks. The latter run only included a subset of stars. These authors present periods up to 10.1 days, with far fewer sub-day periods than \cite{patten1996} find in IC 2391. 
Twenty six rotators from \citet{barnes1999} are also in our membership catalog for IC 2602 and have TESS data.

We also include two additional unpublished periods from archival data.
\citet{tschape2001} reference \prot\ for IC 2391 and IC 2602 stars from a ``Prosser-Stauffer archive'' from 1998, and include some IC 2602 periods in their table 2. 
The reference url for this archive is now broken, and no further information is provided for the provenance of those \prot. 
Copies of the archive files on IC 2391 and IC 2602 were provided by L.\ Rebull (2022, private communication). 
The IC 2391 periods are all from \citet{patten1996}, and the majority of the IC 2602 periods are from \citet{barnes1999}. Additional IC 2602 periods are labeled as ``Apr94 observations by B.\ Patten (Patten, B., Stauffer, J.R., and Prosser, C.F.\ 1996 BAAS 28, 1366)'', which appears to actually be \citet{patten1996-1}, corresponding to a AAS meeting presentation. 
Seven targets from \citet{patten1996-1} also have consistent periods listed from \citet{barnes1999}, while two are new. 
The two unpublished \citet{patten1996-1} periods correspond to stars in our membership catalog for IC~2602, and both have TESS data. 

\citet{irwin2008} present \prot\ for 176 members of NGC 2547 with masses $M_*\ <\  0.9\ \Msun$ (spectral types later than K0). 
As this is the most compact cluster in our sample, it is also the most accessible using standard ground-based imaging. 
Their observations use an 8 month baseline in 2005-2006; no information is given about the frequency of observations. 
Of the rotators from \citet{irwin2008}, 117 are also in our membership catalog for NGC 2547, and 85 have TESS data. 

After accounting for overlaps between catalogs, we have a total of \totalit\ literature periods for stars around 40~Myr. 
The distribution of these periods are shown in Figure~\ref{fig:periodcolor_lit}. 
Thanks to the deep survey carried out by \citet{irwin2008} in NGC~2547, the K and M dwarfs are well-populated. 
There are, however, only \lithigh\ periods for stars with $M\ \gapprox\ 0.9\ \Msun$---not enough to constrain models for stellar angular momentum evolution. 

\section{Derived stellar properties}\label{prop}
To infer stellar parameters for individual cluster members, we utilize an enhanced version of the fitting code, \texttt{MINESweeper} \citep{cargile2020}. The approach this code takes is to derive stellar parameters by determining the posterior distribution of stellar parameters through the modeling of observed spectra and/or photometry, employing the MIST (MESA Isochrones \& Stellar Tracks) stellar evolution models \citep{choi2016}. Recently, an upgraded iteration of the code, referred to as \texttt{uberMS}, has been introduced. A comprehensive description of this upgraded software will be detailed in an upcoming paper (Cargile et al. in prep).
Below, we provide a brief overview of the key enhancements in this update and its relevance to our approach for analyzing stars in our target clusters.

A significant change in \texttt{uberMS} involves the manner in which it samples the posterior distribution of stellar parameters. The code has transitioned to employing Stochastic Variational Inference (SVI), implemented via the probabilistic modeling package NumPyro \citep{phan2019,bingham2019}. This shift from the previously used nested sampling method \citep[dynesty;][]{speagle2020} not only accelerates the inference process but also enhances the code's scalability, particularly for models with a larger number of free parameters. For instance, we can now introduce a "jitter" parameter into the fitting process, effectively accounting for systematic noise-floor effects in the modeled photometry.

Furthermore, we use the most recent version of the MIST models (v2.3; C.~Conroy 2023, private communication)
into \texttt{uberMS}. Optionally, we can also incorporate the influence of cool surface spots on the MIST-predicted stellar parameters using a simple empirical correction calculated from observed spot data given in \citet{berdyugina2005}. This correction adjusts the predicted effective temperature ($T_{eff}$) to account for the presence of a cool spot, and consequently inflating the stellar radius to preserve the total stellar bolometric luminosity.

Given that we possess broadband photometry data for all of our target cluster stars, we employ the photometry-only mode in the fitting procedure with \texttt{uberMS}. 
This entails fitting all available photometric data from Gaia DR3 G, 
$\textrm{G$_{BP}$}$, 
$\textrm{G$_{RP}$}$; 
2MASS J, H, K$_\textrm{s}$; 
and WISE W1, W2 
\citep{gaiadr3,cutri2003cat,WISEmission}.
 
As with any Bayesian modeling framework, we must place prior distributions for all sampled parameters. Specifically, we adopt a uniform prior for distance in conjunction with a normal distribution prior derived from each star's Gaia DR3 parallax. For stellar mass, we employ a Kroupa Initial Mass Function prior \citep{kroupa2001}. Metallicity, alpha-element abundance, and extinction (A$_{V}$) priors are determined based on literature values specific to each cluster (see Table~\ref{tab:lit_metalAv}). The A$_{V}$ prior is a truncated normal distribution which only extends to $3\sigma$ above the cluster mean. 
We test two distinct priors for stellar age in the analysis: 1) a uniform prior ranging from 1 Myr to 1 Gyr, and 2) a normal distribution prior characterized by a mean of 40 Myr and a standard deviation of 15 Myr. The uniform age prior lead to extremely old and unrealistic ages for some K and M dwarfs; we therefore retain only the parameters derived from the $40\pm15$~Myr age prior. 

For each star within our target clusters, posterior distribution functions for all MIST-predicted parameters are computed. From these distributions, we extract maximum a posteriori values, along with associated errors based on the 68\% credible intervals of each marginalized distribution.

\begin{deluxetable}{lcccc}
    \tabletypesize{\footnotesize}
    \tablewidth{0pt}
    \tablecaption{Metallicity and Extinction priors of Target Clusters \label{tab:lit_metalAv}}
    \tablehead{
        \colhead{Cluster} & \colhead{[Fe/H]} & \colhead{$\sigma_{\textrm{[Fe/H]}}$} & 
        \colhead{A$_{V}$} & \colhead{$\sigma_{\textrm{A}}$}
        } 
    \startdata
    Collinder~135 & $-$0.01 & 0.04 & 0.01 & 0.01 \\
    NGC~2451A     & $-$0.08 & 0.06 & 0.01 & 0.01 \\
    NGC~2547      & $-$0.16 & 0.09 & 0.14 & 0.01 \\
    IC~2391       & $-$0.06 & 0.05 & 0.04 & 0.01 \\
    IC~2602       & $-$0.06 & 0.06 & 0.03 & 0.01 
    \enddata
    \tablecomments{Metallicities for IC~2391, IC~2602, NGC~2547, and NGC~2451A are adopted from the Gaia-ESO Survey \citep{bragaglia2022}. For Collinder~135, we use the metallicty measured in the GALAH survey \citep{spina2021}. Extinction values are taken from \citet{cantat2020}.}
\end{deluxetable}


\section{Measuring New Rotation Periods with TESS}\label{prot}

\begin{figure*}[t]\begin{center}
\includegraphics[trim=0cm 0cm 0cm 0cm, clip=True,  width=3.5in]{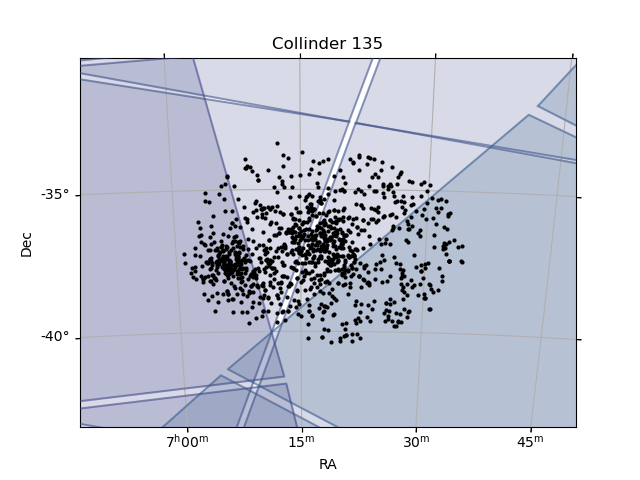}
\includegraphics[trim=0cm 0cm 0cm 0cm, clip=True,  width=3.5in]{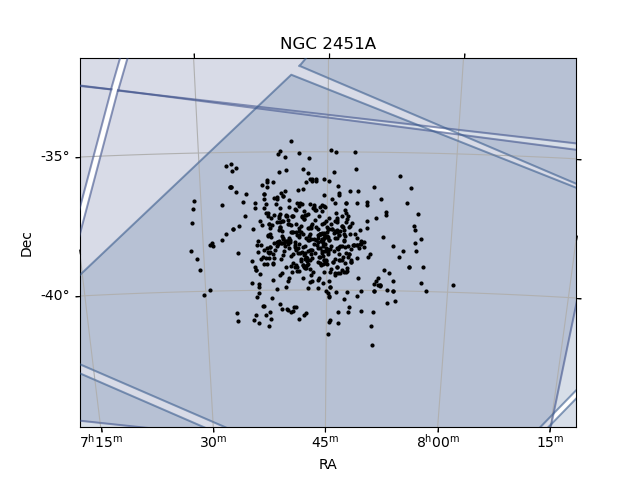}\\
\includegraphics[trim=0cm 0cm 0cm 0cm, clip=True,  width=3.5in]{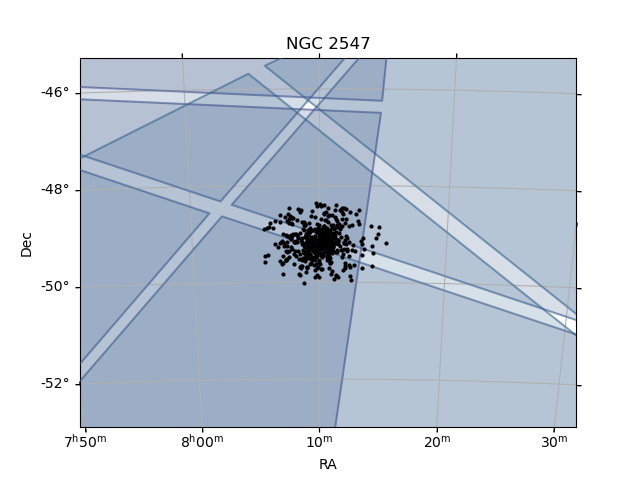}
\includegraphics[trim=0cm 0cm 0cm 0cm, clip=True,  width=3.5in]{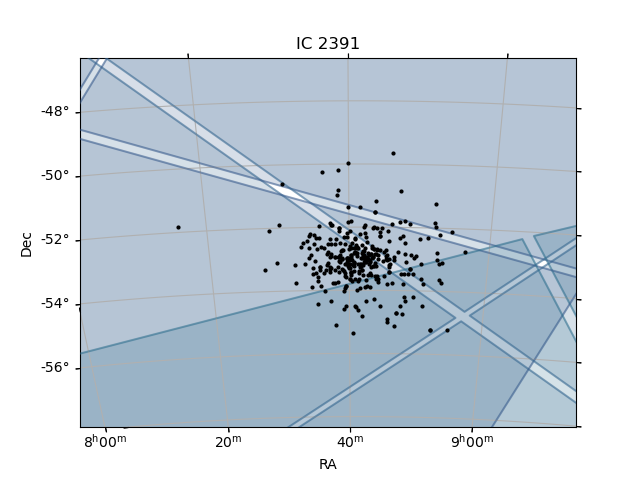}\\
\includegraphics[trim=0cm 0cm 0cm 0cm, clip=True,  width=3.5in]{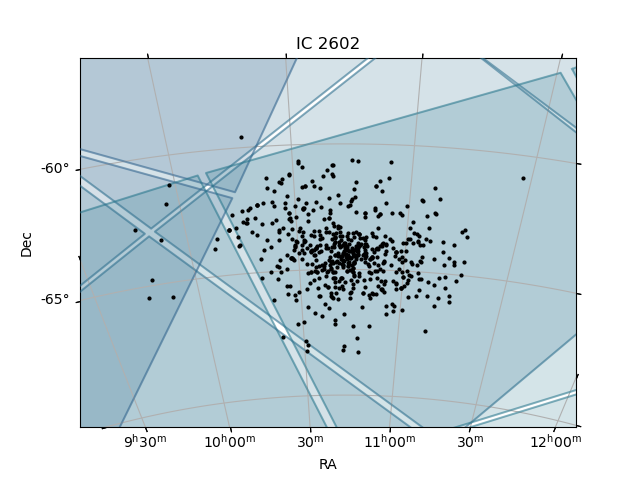}
  \caption{Position of cluster members on the sky, with the TESS camera footprints overlaid. For clarity, only the Cycle 1 sectors that cover each cluster are shown, even if the figure bounds would include additional sectors. Edges of the TESS chips are approximate. Almost all of our targets are in at least one sector, except for a strip of Collinder 135 members that fall in a chip gap in Sector 7. Many cluster members are included in at least two sectors, with a handful of IC 2391 members in three sectors. 
\label{fig:tessfov}
}
\end{center}\end{figure*}

All five of our target clusters were observed by \tess\ in Sectors 6--11, during the first year of the mission. 
Nearly all targets were observed in at least two sectors. 
We analyze the resulting data for \totaltargets\ stars identified in Section~\ref{data}.
The distribution of targets in each cluster is shown in Figure~\ref{fig:tessfov}.
We follow a similar procedure to that used in \citet{douglas2017} and \citet{douglas2019} to analyze \ktwo\ data; we only summarize the method here, and note differences when working with \tess\ data. 

We analyze two sets of community-created full-frame image (FFI) light curves: CDIPS \citep{bouma2019} and QLP \citep{huang2020,huang2020-1}. 
The Cluster Difference Imaging Photometric Survey (CDIPS) uses difference imaging and two different detrending algorithms to extract photometry from the \tess\ FFIs. 
The MIT Quick-Look Pipeline (QLP) uses aperture photometry to extract light curves for stars with \tess\ magnitudes $T\ <\ 13.5$.
We also examined light curves from A PSF-Based Approach to TESS High Quality Data Of Stellar Clusters \citep[PATHOS;][]{nardiello2019}, but jumps and trends meant that these data yielded very few usable periods. 
Both CDIPS and QLP yield light curves without significant long-period trends or jumps in the data, which are important for measuring accurate \prot.

\begin{figure*}[t]
   \centering
   \subfigure[Neither detected period matches the observable repeats in the full light curve; this may be a case of rapid spot evolution or systematics. We set Q~$=2$ as we cannot determine the correct period.]{
   \includegraphics[width=3.25in]{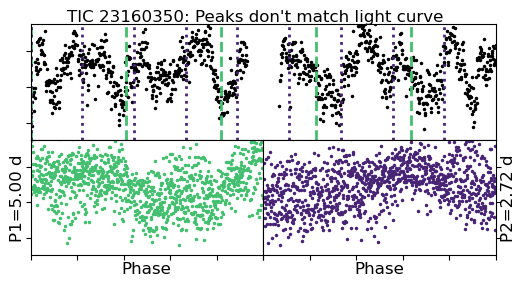}
   }\hfill
   \subfigure[Double-dip structure, periodogram selects half of the likely true period. We select the longer period and set Q~$=0$, with SE?=y.]{
   \includegraphics[width=3.25in]{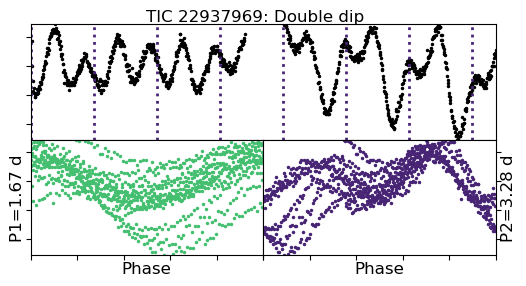}
   }\\
   \subfigure[A systematic trend is detected with high periodogram power; we set Q~$=2$.]{
   \includegraphics[width=3.25in]{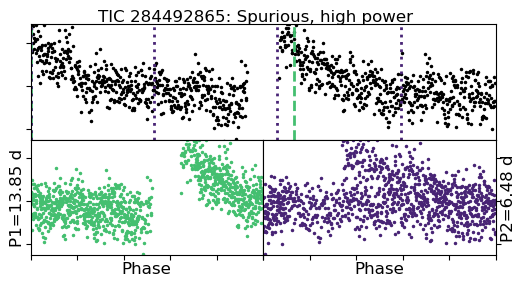}
   }\hfill
   \subfigure[There are two clear periods in the light curve. We set Q~$=0$, and we flag this target as definitely multiperiodic (MP?=y) and therefore a candidate binary.]{
   \includegraphics[width=3.25in]{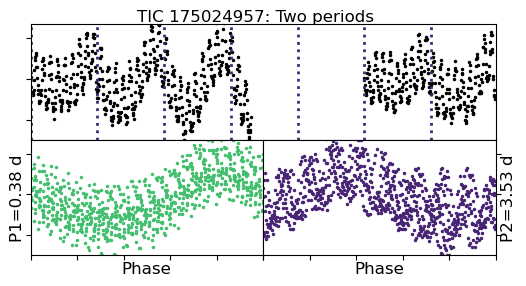}
   }
\caption{Examples of the light curve effects discussed in Section \ref{validation}. 
Vertical lines at intervals of the detected period are overlaid on each full light curve. 
The phase-folded light curves corresponding to the first and second highest periodogram peaks are also shown. 
}\label{fig:pbs}
\vspace{0.6cm}
\end{figure*}

We use the \citet{press1989} FFT-based Lomb-Scargle algorithm\footnote{Implemented as {\it lomb\_scargle\_fast} in the \texttt{gatspy} package; see \url{https://github.com/astroML/gatspy}.} to measure \prot. 
We compute the Lomb-Scargle periodogram power for 3$\times$10$^4$ periods ranging from 0.1 to 70~d (While this is longer than the typical sector length of $\sim$27~d, in practice we never detect periods longer than $\sim$20~d).
We also compute minimum significance thresholds for the periodogram peaks using bootstrap re-sampling, and only consider a peak to be significant if its power is greater than the minimum significance threshold for that light curve.
We take the highest significant peak as our default \prot\ value.

Due to the sector length and mid-sector data downlink, it is challenging to confidently identify periods longer than $\approx12$~d in the \tess\ data. 
However, all of our target stars should be at the shortest \prot\ stage of their lifetimes. 
In h Per (13 Myr), \citet{moraux2013} find a maximum period of $\approx\ 16$~d, with most stars having $P_{rot}\ \lesssim\ 9$~d. 
In the Pleiades (125 Myr), nearly all stars have $P_{rot}\ \lesssim\ 10$~d, with a handful of early M dwarfs extending to $P_{rot}\ \lesssim\ 20$~d \citep{rebull2016}.
These two clusters bracket the ZAMS: PMS contraction will cause stars to spin up relative to the h Per values, before spinning down again to reach Pleiades values. 
Even in the case of minimal spin-up, we can still assume that most PMS stars have $P_{rot}\ \lesssim\ 10$~d.
Therefore, a simple Lomb-Scargle analysis is sufficient to identify \prot\ in these clusters. 

We may miss some longer periods in the M dwarf regime, but \citet{irwin2008} measure ground-based periods for K and M dwarfs in NGC~2457 using a longer baseline.
Combining our shorter \tess\ periods with this ground-based data is sufficient to assess rotational evolution for low-mass stars at this age. 

We analyze each sector and each aperture/detrending method individually.
Because we are searching for shorter periods, stitching sectors together is more likely to cause jumps and long period trends, instead of increasing our sensitivity. 
For CDIPS, we examine light curves detrended using both the principle component analysis (PCA) and trend-filtering algorithm (TFA), through three different apertures; in almost all cases, the PCA light curves provide a cleaner signal. 
For QLP, we examine only the SAP light curves, since other light curves have been detrended to the point of excluding some stellar signals. 
Therefore, depending on the star, we have anywhere from one to twenty one light curves and associated periodograms.

\begin{figure*}[tp]\begin{center}
\includegraphics[width=\textwidth]{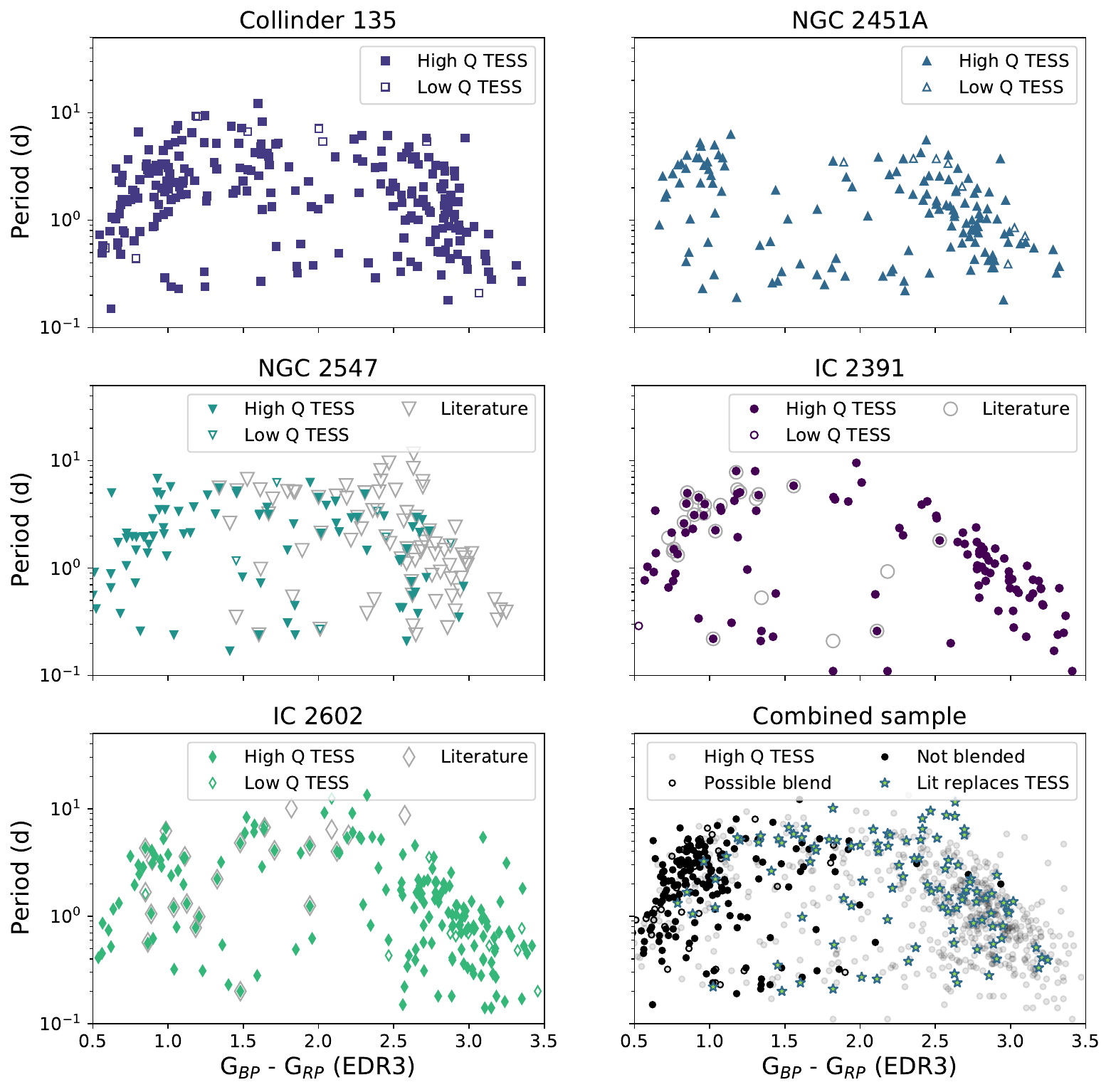}
  \caption{TESS periods for all of our target stars. In the individual cluster panels, colored solid (open) symbols represent high-(low-)confidence TESS periods with $Q1=0$ ($Q1=1$). Open grey symbols in the individual panels indicate literature periods for any star in our membership catalog.
  In the lower right panel, we show all TESS periods with $Q1=0$. Unblended or potentially blended stars are shown in black closed or open dots, respectively; likely blends are shown in grey. 
  Where we have \prot\ from both TESS and the literature, but the TESS \prot\ is lower quality or definitely blended, we replace it with the literature period (colored star symbols). When we compare measured \prot\ to theoretical models in Section~\ref{tausq}, we focus on the sample of unblended TESS stars plus literature replacements. 
\label{fig:results}
}
\end{center}\end{figure*}

\subsection{Period Validation}\label{validation}

We visually inspect the light curves for each target.
We use the phase-folded light curve to confirm that the detected \prot\ appears astrophysical and not instrumental.
We also plot vertical lines on the full light curve at intervals corresponding to the highest two periodogram peaks, and ensure that similar light curve features repeat at every line. 
Clearly spurious detections are flagged as Q~$=2$, and questionable detections as Q~$=1$.
In contrast to \citet{douglas2017,douglas2019}, we are more strict about accepting periods, and place a much higher proportion of stars into the Q~$=2$ category. 
A Q~$=3$ flag indicates that there were no significant periodogram peaks.
Figure~\ref{fig:pbs} shows examples of various light curve features, and describes how we flag them.

During our visual inspection, we also flag stars with spot evolution and/or multiple periods apparent in the light curve. 
In Table~\ref{tab:all_stars}, we flag spot evolution (SE?) and multiperiodic stars (MP?) as y, m, or n for ``yes'', ``maybe'', and ``no'', respectively. 
In some cases, such as panel (d) in Figure~\ref{fig:pbs}, both periods are detected in the periodogram. 
In other cases, a possible second period visible by eye, but may not be detected at high significance. 
Therefore a star may be flagged as MP?=y, without a second period reported.
Panel (b) of Figure~\ref{fig:pbs} shows an example of a star with clear spot evolution (SE?=y).

\subsection{Contamination within TESS pixels}\label{contamination}

Finally, we use Gaia astrometry to identify possible blends or contamination from nearby stars. 
In previous work, we relied on visual inspection of the \textit{K2} pixel stamp and existing imaging data to identify possible blends. 
Given TESS's large pixel size, there is a much higher chance that we will miss an overlapping star in our visual inspection. 
While we include contamination ratio (Rcont) values from the TIC \citep{tic2019} in Table~\ref{tab:all_stars}, values are not available for all of our target stars. We therefore choose to perform our own blending analysis based on Gaia data, and do not take Rcont into account.

We adjust our final period catalog based on automated analysis of the Gaia EDR3 data. 
We perform two check for Gaia sources near each target, and the results are flagged as blends (Bl?) in Table~\ref{tab:all_stars}.
A definite blend (y for ``yes'') indicates a neighbor within 30 arcseconds ($\sim$1 pixel) with a TESS magnitude brighter than $T_\textrm{target}-3$. 
A total of 1029 targets, including 673 with measured \prot, are likely contaiminated and have Bl?=y.
Injection tests indicate that signals from stars more than three magnitudes fainter than our target are undetectable in $>90\%$ of cases. 
A possible blend (m for ``maybe'') indicates a potential contaminant within 1 arcminute of the target and with  $T\le12$. 
Seventy four targets, including 45 with meausured \prot, are potentially blended with a neighboring star. 
Stars without potentially contaminating neighbors in Gaia are flagged with n for ``no''. 
A total of 307 targets, including 216 with periods, are not blended with other stars. 
As can be seen in the lower right panel of Figure~\ref{fig:results}, the majority of blended stars are faint M dwarfs.

If the neighboring targets have different TESS periods, then we allow both periods in the final catalog. 
If two close targets have a period that matches within 5\%, we assess the relative brightness and periodogram powers of each source. 
If the neighboring source is at least 1 magnitude brighter or has a periodogram power at least 0.1 higher, then we assign the period to that source. 
This clear signal contamination only occurs for four stars.

\startlongtable
\begin{deluxetable*}{lll}
    \tabletypesize{\footnotesize}
    \tablewidth{0pt}
    \tablecaption{Description of columns for the full catalog \label{tab:all_stars}}
    \tablehead{
        \colhead{No.} & \colhead{Name} & \colhead{Description}
    } 
    \startdata
1 & TIC & TESS Input Catalog identifier \\
2 & GAIAEDR3\_ID & Gaia EDR3 identifier \\
3 & GAIAEDR3\_RA & Right Ascension in decimal degrees (J2015.5) \\
4 & GAIAEDR3\_DEC & Declination in decimal degrees (J2015.5) \\
5 & GAIAEDR3\_PMRA & Gaia EDR3 proper motion in right ascension \\
6 & GAIAEDR3\_PMDEC & Gaia EDR3 proper motion in declination \\
7 & GAIAEDR3\_PARALLAX & Gaia EDR3 parallax\\
8 & GAIAEDR3\_PARALLAX\_CORRECTED & Gaia EDR3 parallax, with zero-point correction \\
9 & GAIAEDR3\_RUWE & Gaia EDR3 Renormalized Unit Weight Error \\
10 & GAIAEDR3\_G & Gaia EDR3 G-band magnitude \\
11 & GAIAEDR3\_G\_ERR & Gaia EDR3 G-band magnitude \\
12 & GAIAEDR3\_G\_CORRECTED & Corrected Gaia EDR3 G-band magnitude (equals the Gaia DR3 G-band magnitude) \\
13 & GAIAEDR3\_BP & Gaia EDR3 blue magnitude \\
14 & GAIAEDR3\_BP\_ERR & Error on Gaia EDR3 blue magnitude \\
15 & GAIAEDR3\_RP & Gaia EDR3 red magnitude \\
16 & GAIAEDR3\_RP\_ERR & Error on Gaia EDR3 red magnitude \\
17 & TIC\_Tmag & TESS magnitude from the TIC \\
18 & TIC\_Rcont & Contamination ratio from the TIC, if available\\
19 & TIC\_Ncont & Number of stars used to compute TIC\_Rcont\\
20 & HDBscan\_MemProb & Membership probability from HDBScan analysis \\
21 & HDBscan\_Cluster & Clump identifier from HDBScan analysis \\
22 & HDBscan\_Stability & Stability of the clump from HDBScan analysis \\
23 & MemBool & Whether the target should be considered a member from HDBScan analysis (0=no, 1=yes) \\
24 & angDist\_GES & Angular separation between Gaia EDR3 position and GES position\\
25 & GES\_Target & Gaia-ESO Survey (GES) identifier \\
26 & GES\_Cluster & Cluster name from GES \\
27 & GES\_MemProb & Membership probability from GES ($>=$0.9 and $<=$1.0 indicates a likely member) \\
28 & angDist\_Cantat-Gaudin & Angular separation between Gaia EDR3 position and Cantat-Gaudin position\\
29 & CG\_MemProb & Membership probability from Cantat-Gaudin ($>=$0.7 and $<=$1.0 indicates a likely member\\
30 & CG\_Cluster & Cluster name from Cantat-Gaudin\\
31 & av & extinction in V-band from MINESweeper analysis \\
32 & av\_err & error on extinction in V-band from MINESweeper analysis  \\
33 & dist & distance in pc from MINESweeper analysis \\
34 & dist\_err & error on distance in pc from MINESweeper analysis \\
35 & log(Age) & age from MINESweeper analysis  \\
36 & log(Age)\_err & error on age from MINESweeper analysis  \\
37 & Mass & stellar mass (solar masses) from MINESweeper analysis  \\
38 & Mass\_err & error on stellar mass (solar masses) from MINESweeper analysis \\
39 & log(Teff) & effective temperature (K) from MINESweeper analysis \\
40 & log(Teff)\_err & error on effective temperature (K) from MINESweeper analysis \\
41 & Prot1 & Rotational period, first periodogram peak\\
42 & Pw1 & Power, first periodogram peak \\
43 & Q1 & Quality flag, first periodogram peak (1)\\
44 & Sig & Minimum significance threshold for periodogram peaks \\
45 & Prot2 & Rotational Period, second periodogram peak, if any\\
46 & Pw2 & Power, second periodogram peak, if any\\
47 & Q2 & Quality flag, second periodogram peak, if any (1)\\
48 & MP? & Multiple rotational periods? (M=Maybe)\\
49 & SE? & Spot evolution? (M=Maybe)\\
50 & Bl? & Blend? (M=Maybe)\\
51 & ClosestNeighborSep & Angular distance to closest neighbor in Gaia EDR3 \\
52 & ClosestNeighborMagDiff & Magnitude difference with closest neighbor in Gaia EDR3 \\
53 & BrightestNeighborSep & Angular distance to the brightest star within 1 arcmin in Gaia EDR3 \\
54 & BrightestNeighborMagDiff & Magnitude difference with the brightest star within 1 arcmin in Gaia EDR3 \\
55 & LitPeriod & Rotational period from the literature, if any \\
56 & LitSource & Source of rotational period from the literature, if any (2)\\
57 & Cluster & Cluster this star is a member of \\
58 & to\_plot & Whether this star is plotted in the paper (1=yes) 
    \enddata
\end{deluxetable*}


\section{New TESS Periods for ZAMS clusters}\label{res}

In total, we measure \tessperiods\ rotation periods for stars across all five clusters from TESS data. 
\newperiods\ of these are new values, and \tesslit\ stars also have literature \prot. 
Our new \prot\ are shown in Figure~\ref{fig:results}, and comparison with the literature is discussed in Section~\ref{lit_comp}.

All five clusters show similar morphology in the period-color plane.
The downturn at the higher-mass end represents stars which have reached the main sequence and are beginning to converge onto the slow-rotator sequence. 
Stars slightly lower-mass than the Sun are approaching the main sequence, and still show a wider range in periods. 
The ``gap'' between the fast and slow sequences \citep[e.g.,][]{barnes2003} is clearest in Collinder~135 and IC~2391, and in the cleaned TESS plus literature sample in the lower right panel of Figure~\ref{fig:results}.
Finally, the reddest/lowest-mass stars are still contracting and spinning up to shorter \prot; this accounts for the downturn at the low-mass end.

\subsection{Comparison to literature periods}\label{lit_comp}

\begin{figure*}[t]
\centerline{\includegraphics[width=6.5in]{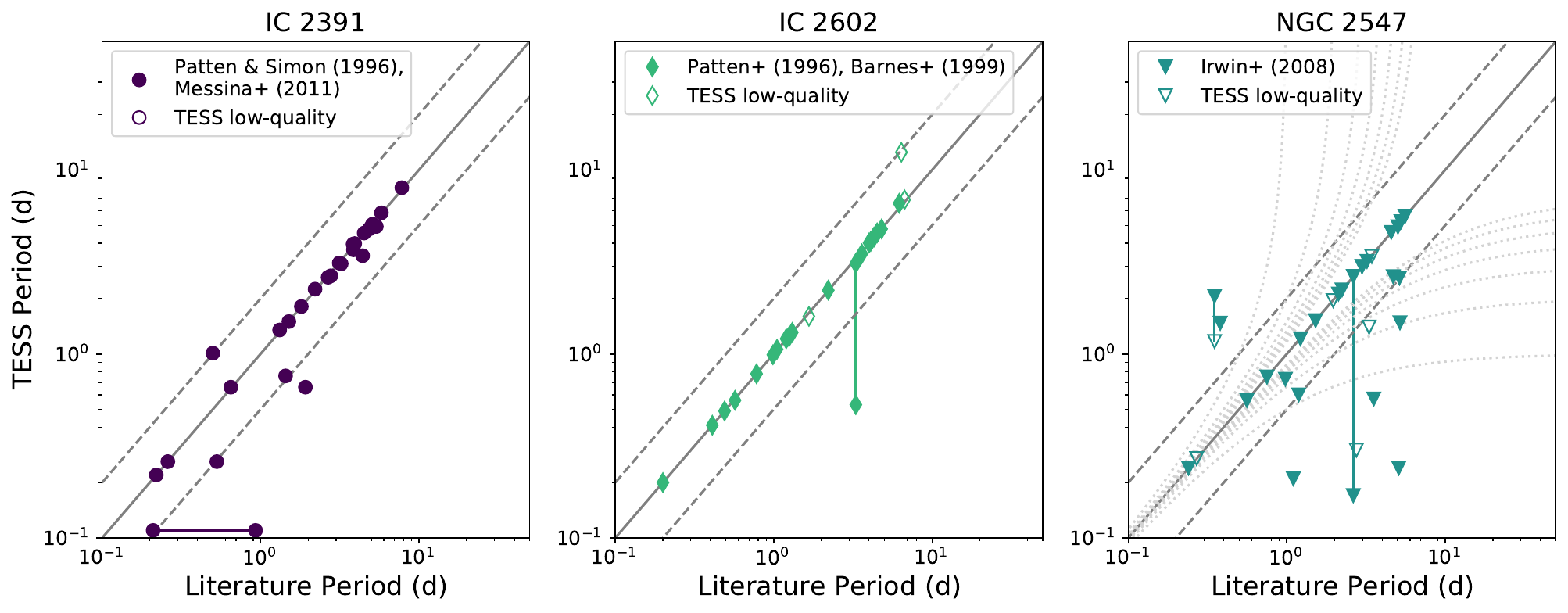}}
\caption{Literature \prot\ for IC 2391, NGC 2547, and IC 2602 compared to TESS results. The solid grey line shows a one-to-one match; dashed lines indicate double- or half-period harmonics. A horizontal line in the IC~2391 panel connects a resolved binary discussed in Section~\ref{lit_comp}. In IC 2602 and IC 2391, our TESS results are largely consistent with the literature.  In the NGC 2547 panel, dotted grey lines indicate potential aliases in the ground-based data, and vertical lines connect multiple TESS periods measured for the same star. NGC 2547 is the most crowded cluster in our sample, and these discrepancies with \citet{irwin2008} are likely due to blending in TESS (see Sections~\ref{contamination} and \ref{lit_comp}).
}
\label{fig:lit_comparison}
\end{figure*}

\begin{figure*}[t]
\centerline{\includegraphics[width=6.5in]{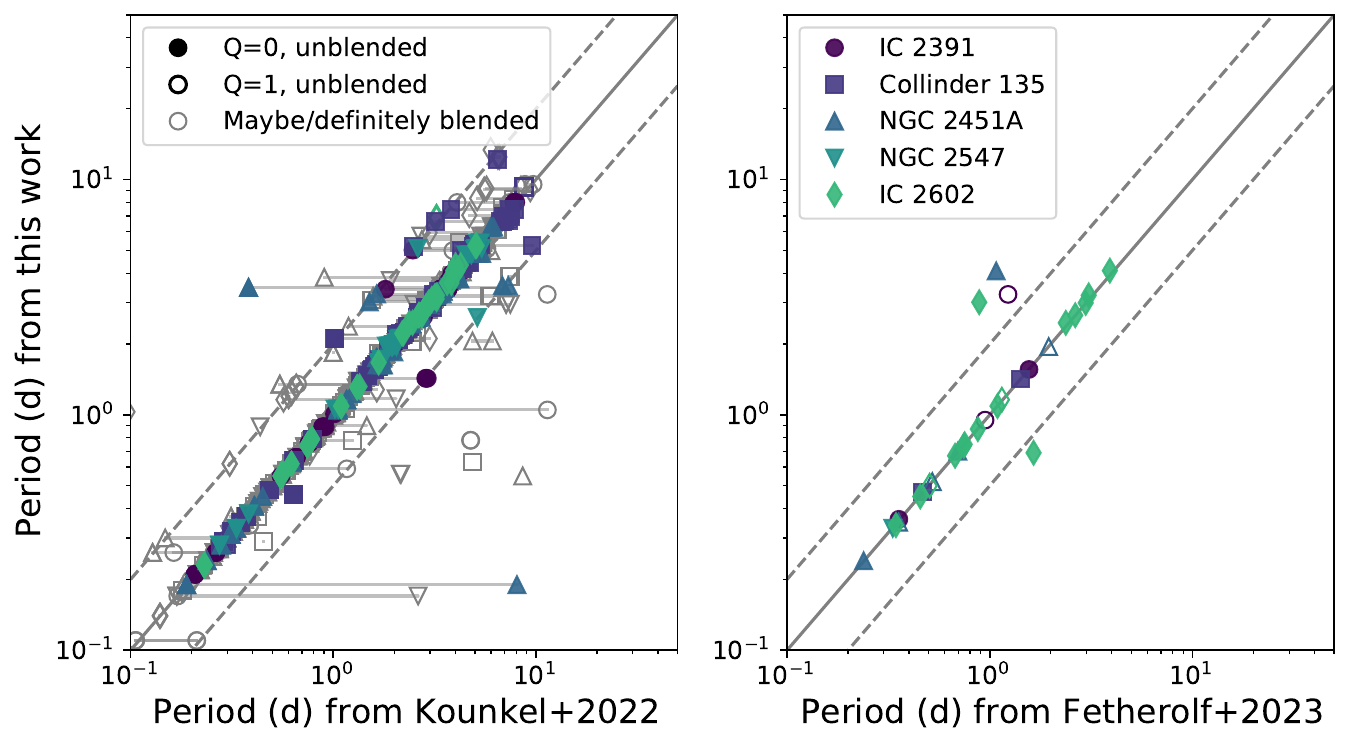}}
\caption{Literature \prot\ from \citet{kounkel2022} and \citet{fetherolf2023} compared to our results. Colors/shapes are as in Figure~\ref{fig:lit_comparison}, and grey outlines indicate stars that are likely blended. Horizontal lines in the left plot show where \citet{kounkel2022} measure different periods in different TESS sectors. The solid grey line shows a one-to-one match; dashed lines indicate double- or half-period harmonics. Overall, our results are in good agreement with these other surveys. 
}
\label{fig:lit_comparison2}
\end{figure*}

As discussed in Section~\ref{lit}, there have been four rotation studies targeting three of our clusters in the past.
\citet{patten1996} include 15 F8-M3 stars in IC~2391; \citet{patten1996-1} and \citet{barnes1999} include 28 F-M stars in IC~2602; and \cite{irwin2008} include 85 K-M stars in NGC~2547.
These literature periods are compared with the TESS results in Figure~\ref{fig:lit_comparison}. 

In the two clusters with literature data for F-G stars, we find few discrepancies between our periods and the TESS results. 
We have no mismatches with the \citet{patten1996-1} and \citet{barnes1999} results for IC~2602.
We do find several discrepancies with the \citet{patten1996} results for IC~2391; the VXR names given here are the corresponding names in that paper.

Three discrepancies in IC~2391 are consistent with half- or double-period harmonics. 
The TESS \prot\ for TIC~93833881/ VXR 35a (0.26~d) is half of the Patten \& Simon \prot\ (0.527~d).
The TESS \prot\ for TIC~94184691/VXR 60b (0.11~d) is also half of the Patten \& Simon value (0.212~d).
In both cases, neither TESS nor the phase-folded Patten \& Simon plots show strong evidence for a double-dipped structure, although there is a short but significant peak in the TESS data at the longer \prot. 
The TESS light curve for TIC~94185535/VXR 62a shows a clear double-dip signature and \prot$=1.01$~d, while the Patten et al.\ phase-folded light curve only spans half a cycle for their measured \prot$=0.50$~d; this suggests that the literature result is likely a half-period harmonic of the true period. 
These types of harmonics are unsurprising, given spot evolution and the sampling issues of ground-based surveys. 

TIC~812594503/VXR 60a has very different periods in TESS (0.11~d) and Patten \& Simon (0.93d). 
As suggested by their literature names, TIC~812594503/ VXR~60a and TIC~94184691/ VXR~60b are a binary system. 
The system is barely resolved in 2MASS or digital sky survey (DSS) imaging accessed through Simbad. 
They are therefore completely blended in TESS, and the TESS light curve is dominated by a 0.11~d signal. 
There are also significant peaks in the TESS data at $\sim$0.22~d and 1.91~d, though we did not flag this as a believable period in our visual validation step. 
The Patten \& Simon 0.93~d phase-folded light curve for TIC~812594503/VXR 60a is noisier than for their other detections; this probably represents an alias or perhaps half of the 1.91~d period visible in TESS. 
For consistency, we do not remove this target from our sample, since we do not have clear external \prot\ checks for most targets. 
We leave the 0.11~d TESS period assigned to both targets, even though it more likely represents a half-period harmonic for TIC~94184691.

We find the most discrepancies with NGC~2547 from \citet{irwin2008}. 
Since these authors focused only on K and M dwarfs, the comparison stars in this cluster are much fainter than in IC~2391 or IC~2602. 
Furthermore, NGC~2547 is the most distant cluster in our sample, and the most densely concentrated on the sky: blending is therefore a much more significant issue. 
We attribute discrepancies between TESS \prot\ and values from \citet{irwin2008} to mismatches due to blending. 

We also compare our results to two larger studies of rotation in TESS (Figure~\ref{fig:lit_comparison2}) 
A total of 512 rotators in our sample overlap with \citet{kounkel2022}; the majority of their measurements are consistent with ours, or with half/double-period harmonics. Of 135 overlapping stars where we flag the TESS period with Q=0 and our Gaia search shows no possible contaminants, all are consistent with the \citet{kounkel2022} values in at least one sector.
Twenty-nine of our rotators overlap with \citet{fetherolf2023}: four of these show discrepant measurements compared to our results, and only three of our detections have Q=0. TIC 173512656 and TIC 460796730 show stronger variability at the longer period we selected (4~d and 3~d, respectively), with the shorter variability visible but not dominant. These are likely cases where rapidly evolving spot configurations were interpreted as the rotation period by these authors' analysis methods. TIC 389988925 shows no evidence for a 1.6d period in the QLP or CDIPS light curves, though there are multiple longer peaks in the periodogram which we discarded as likely systematics. \citet{fetherolf2023} use only 2-minute cadence PDCSAP light curves, while we use only FFI light curves, which may also explain the different periods detected. 
Overall, however, our periods are consistent with other analyses of TESS data.

Overall, our new \tess\ \prot\ measurements are consistent with existing literature results for F, G, and early K stars. 
This gives us confidence in our detections for stars at these masses.
Consistent with our prior \ktwo\ surveys, it also highlights the ability of ground-based surveys to measure accurate \prot\ for reasonably bright and unblended targets. 
There is some doubt about our measurements where multiple stars fall within a TESS pixel, which particularly impacts late K and early M stars; we therefore remove all potentially blended stars from further analysis in the next section. 

\section{Comparison to models of angular momentum evolution}\label{tausq}

Our dataset significantly increases the number of measured rotation periods for stars with masses near solar and at ages near their ZAMS age.
The larger catalog enables better characterization of the \prot\ distribution during a time when these stars' rotation rates are observed to be near their maximum value.
Thus, the dataset should be particularly useful for understanding the angular-momentum evolution of near-solar-mass stars during and up to the end of the pre-main-sequence (contraction) phase.
Here we compare the observations to a few simplified models, primarily to demonstrate the usefulness of the dataset; a more detailed comparison is left for future work.
For this comparison, we use the dataset that includes TESS periods only for unblended stars, replacing low-quality TESS detections and blended stars with literature periods when available.
In this case, rotation periods for stars at the low-mass end primarily come from NGC~2547, measured by \citet{irwin2008}.

\subsection{Description of spin-evolution models}

\begin{figure}[!t]
\centerline{\includegraphics[width=\columnwidth]{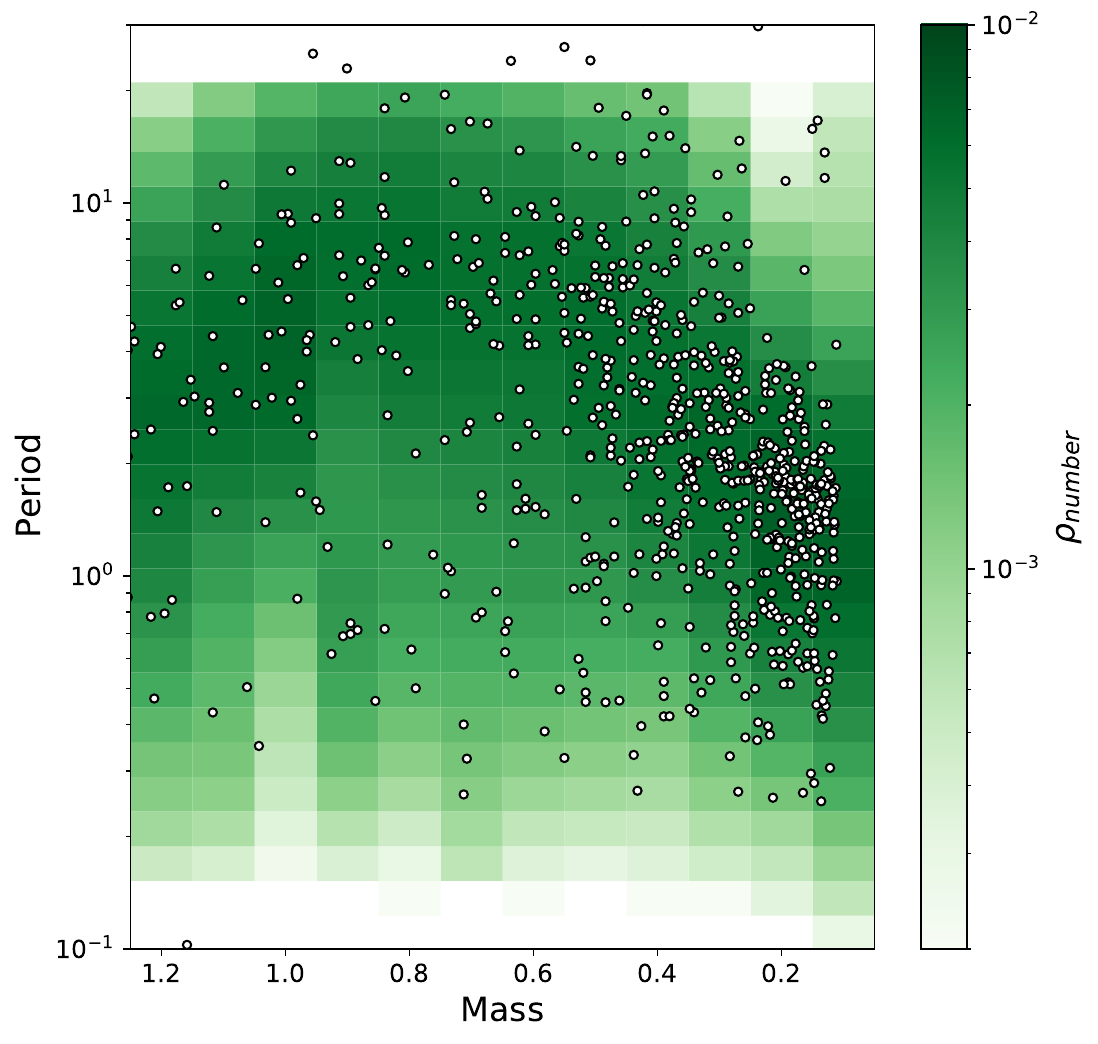}}
\caption{Observed rotation periods for Upper Sco \citep[dots;][]{rebull2018} overplotted on a KDE approximation to the data, which spans 0.1 to 20~d. To reduce the impact of individual outliers on the modeled period-mass distribution, we use the KDE to initialize the spin-evolution models, instead of the Upper Sco stars themselves. 
}
\label{fig:upsco_kde}
\end{figure}

Theoretically, the rotation rate of the visible surface of a star can change with time, due to (a) changes in the moment of inertia of the star (e.g., caused by contraction or redistribution of mass in their interior), (b) changes in the distribution of angular momentum in the interior, and (c) changes in the total angular momentum content (e.g., caused by external torques arising from stellar winds).
We summarize how we account for each of these below; the models are described in more detail in \citet{breimann2021}.

(a) To track the changes in stellar structure, we use structural evolution models of \citet{baraffe1998}.
Using pre-computed structural evolution models in this way assumes that rotation has a negligible effect on the structural evolution, which should be true for all stars but those rotating at a large fraction of the breakup rate (e.g., for solar-mass stars with \prot\ $\la0.2$ days, and shorter limiting periods for lower mass stars).

(b) We assume the redistribution of angular momentum in the interior is instantaneous, such that stars are assumed to rotate as solid bodies.
This assumption is for simplicity and also is useful as a representation of the limiting case of efficient internal transport (because the true transport rate and mechanism physics are still uncertain; e.g., \citealp{amard2016}).

(c) For modeling the external torque on the star, we compute three different scenarios.
The first case assumes there is no torque, so that stars conserve their angular momentum as they evolve (i.e., they spin up during pre-main-sequence contraction and then maintain approximately constant rotation rate after their ZAMS).
This ``zero-torque'' model also serves as a control model, to disentangle how the ZAMS behavior is affected by contraction vs.\ magnetic braking.

Then we test two different stellar-wind-torque prescriptions, to demonstrate how the new dataset can be used to discriminate between models.
We use the torque formulation presented in \citet{breimann2021}, which assumes the magnetic properties of a star (a combination of the mass-loss rate and magnetic field strength) follows a broken power-law relationship with Rossby number.  
The formulation has three free (fitting) parameters.  
$p_{\rm s}$ and $p$ are the power-law exponents on Rossby number for the saturated and unsaturated regime, respectively. 
$k_{\rm s}$ provides a normalization for the saturated regime that effectively determines at which Rossby number the two power-laws meet.
$\beta$ describes the magneto-centrifugal acceleration; it affects the torque very slightly and only for faster rotators (in the saturated regime), and setting $\beta=1$ neglects this physical effect.

One stellar-wind-torque model, hereafter referred to as the ``classical'' torque, is given by $k_{\rm s}=100$, $p_{\rm s}=0$, $p=2$, and $\beta=1$ (which also gives a torque identical to that presented by \citealp{matt2015}).
The other stellar-wind-torque model, hereafter referred to as the ``standard'' torque, is given by $k_{\rm s}=450$, $p_{\rm s}=0.2$, $p=2$, and $\beta$ computed by equation 5 in \citet{breimann2021}.
The classical- and standard-torque models approximately represent the two extremes in the range of models explored by \citet{breimann2021} (see, e.g., figure 10 in that work).  
The difference between the two torque formulations is in the behavior of the torque for rapidly rotating stars (in the so-called ``saturated'' regime).
The standard-torque model was tuned to better reproduce the distributions of rapidly rotating stars observed in the Pleiades and Praesepe star clusters. 
At the later stages of spin-down (in the ``unsaturated'' regime), the two torques are identical.
The spin-evolution models predict the rotation rate, as a function of time, for any individual star with a specified mass, initial rotation rate, and external torque prescription.

\subsection{Predicting period-mass distributions and calculating goodness-of-fit}\label{gof}

In order to quantitatively compare the full observed mass-period distribution to models, we compute the goodness-of-fit parameter \tausq.
The \tausq\ method was first developed by \citet{naylor2006} for isochrone fitting; it is analogous to $\chi^2$, but modified to work with two-dimensional probability distributions. 
Smaller \tausq\ values imply a better fit, but the absolute \tausq\ numbers depend on the number of observed data points. 
Therefore, \tausq\ fitting is primarily useful when comparing multiple models to the same dataset.

\begin{figure*}[t]
\centerline{\includegraphics[width=\textwidth]{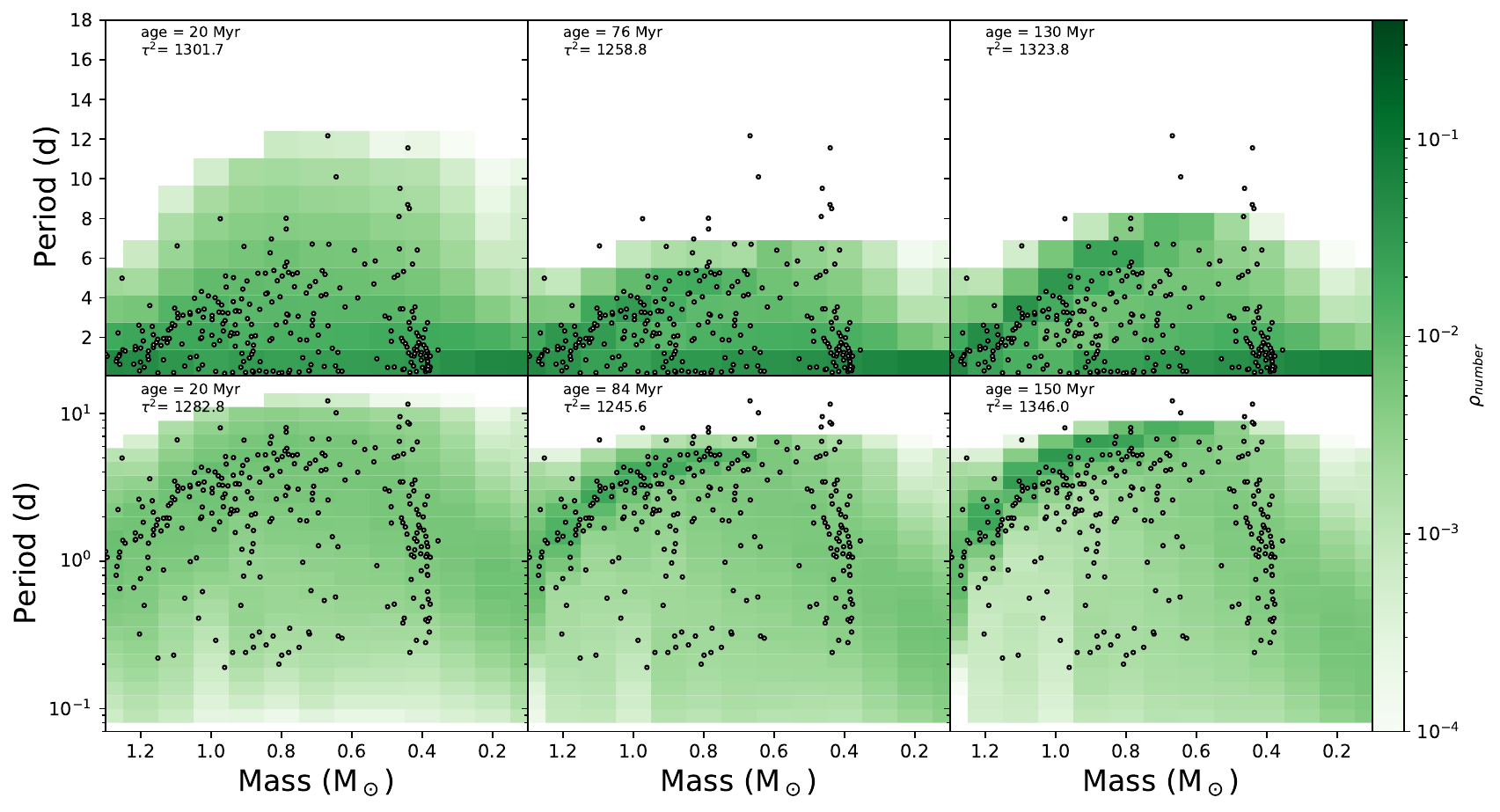}}
\caption{Tau-squared shown at different ages for the standard model, with data for ZAMS clusters overlaid. 
The top and bottom row show the periods in linear and log scale, respectively.
The left, middle, and right columns show the models at three different ages, as indicated in each panel.
The fits in the center panels have the lowest \tausq\ values, and those models have the best visual match to the observed data. 
}
\label{fig:tausq_panel}
\end{figure*}

The \tausq\ method compares an observed dataset to a probability density distribution predicted by a model in two-dimensional space---in this, case period-mass space.
In order to use the spin-evolution models to predict a period-mass distribution, we follow the method described in \citet{breimann2021} and use the period-mass distribution of Upper Sco ($\sim$8~Myr) as an initial condition.
We initialize a grid of stars in period-mass space and compute their rotational evolution according to the three different torque prescriptions.
The model grid is comprised of 13 discrete mass bins, from 0.1 to 1.3 solar masses, with a spacing of 0.1 solar masses. 
Each mass bin contains 500 initial periods, evenly spaced in the logarithm of period from 0.1 to 20 days.

To represent the distribution in Upper Sco, each model star is assigned a weighting value, based on its initial position in period-mass space. 
The weighting values are derived from a kernel-density-estimate (KDE) fit to the log-period distribution observed in Upper Sco, binned in mass to match the model grid.
The KDE approximates each \prot\ in Upper Sco as a Gaussian (with a width in log-period of 0.3), with the sum of all giving the probability density function used for weighting model stars in each mass bin.

For the Upper Sco dataset, we adopt an age of 8~Myr, \prot\ from \citet{rebull2018}, and mass determinations from \citet{breimann2021}. 
Figure~\ref{fig:upsco_kde} visualizes the weighted distribution of model stars with individual observations overlaid.
The regions with more observed data points generally correspond to regions of higher model density, by design.
Outliers, however, are either smoothly incorporated into the overall distribution or excluded if they fall outside the 0.1 to 20 day window. 
This smoothed approximation of the observed data ensures the randomness of individual stars in Upper Sco does not impact the fit to other clusters.

After evolving the model stars, at each age, the predicted model probability density is computed in period bins (30 bins between 0.08 and 40 days), by summing the weighting value of each model star in each bin.
Finally, the \tausq\ statistic is computed by summing the log of the predicted probability density of the models over each observed star's location in period-mass space (for formulations and further details, see \citealp{breimann2021}).
We carry out fits in both log and linear period space. 

\subsection{Model-data comparison}

Figure~\ref{fig:tausq_panel} shows the evolution of the model distribution (color scale), predicted by the standard-torque model, compared to our new dataset (data points).  
Between the first two columns, the model distribution shifts toward shorter rotation periods, due to the spin-up of stars as they contract toward the ZAMS (which occurs at an age of $\sim$40~Myr, for solar-mass stars).
By the age of the middle panels, the model distributions have started forming a ``slow-rotator sequence'', visible as a mass-dependent overdense region among the slower rotators in the mass range of 0.6--1.2 solar masses.
In subsequent evolution, stars in the slow-rotator sequence continue to spin down, and the structure in the model distributions becomes more pronounced.

In comparing to the observed data, the middle column shows the models at the age of the best fit to the data, most notably showing the best coincidence of the location of the slow-rotator sequence visible in both the model and data.
At the younger ages (left column), the model distribution generally overlaps with the observed cluster data, but the model predicts many more slow rotators than observed and has less structure than in the observed distribution.
And at the older ages (right column), the modeled slow-rotator sequence has become narrower and evolved to a slower range of rotation periods than exhibited by the observed dataset.
The \tausq\ values in each panel corroborate these visual inspections: the best-fitting models are at an age of $\approx80$~Myr where the slow-rotator sequence of the model and data overlap. 

Figure \ref{fig:tausq_tracks} shows the evolution of \tausq\ with age for all three torque models. 
The location of global minima represent the age at which the models best fit the data. 
The shape of these curves is similar for both log and linear fits, and our descriptions below apply to both. 
For the first $\approx20$~Myr, all three models produce roughly the same \tausq\ values, indicating that contraction is the only factor in the angular momentum evolution so far. 
From that point, the models diverge. Stars in the zero-torque model continue to spin-up as they contract, but this actually moves them to faster periods than observed in our dataset. 
In contrast, the two models with external torques produce an increasingly better fit---indicating that stellar winds likely affect angular momentum evolution well before stars reach the ZAMS.

Both the classical- and standard-torque models produce a local minimum in \tausq\ at $\approx30$~Myr, since the model stars overlap with the observations during their spin-up. 
As discussed above, however, the model stars retain their initially broad distribution at this age, and do not replicate the partially converged sequence observed in the ZAMS data. 
The classical and standard models then reach a local maximum in \tausq\ as the model stars reach their fastest \prot, having passed most of the observed stars.

After $\approx30$~Myr, the effects of magnetic braking become significant, and both the classical and standard models produce a global minimum \tausq\ value. 
The standard model reaches a lower global value of \tausq\ (indicating a better fit) than the classical model, and it does so at younger best-fit ages: 82~Myr (linear fit) and 84~Myr (log fit).
These best-fit ages correspond to the middle column of Figure~\ref{fig:tausq_panel}. 
The classical model yields older best-fit ages of 130~Myr (linear fit) and 126~Myr (log fit), with a shallower minimum in the \tausq\ curve. 
This behavior indicates that the standard model reproduces the observed period-mass distribution (particularly the slow-rotator sequence) more quickly and accurately than the classical model.

\subsection{Uncertainties on the \tausq\ fits}\label{unc}

We follow the method of \citet{naylor2006} to derive uncertainties on the best-fit ages from each model. 
We determine the number of observed stars in each mass bin, and randomly draw the same number of stars from the model probability distribution to produce a synthetic dataset. 
We then fit the same model to this synthetic dataset, and determine the best-fit age. 
We produce 100 synthetic datasets from each model, and take our confidence interval from the age range containing 67\% of the best-fit ages. 
This method estimates the inherent uncertainty in the models themselves, and the confidence intervals are listed in Table~\ref{tab:res}.

We also resample the observed stellar masses to derive uncertainties on the model fits, since the mass uncertainties dominate the observed period-mass distributions.
We redraw each star's mass from within the probability distribution inferred in Section~\ref{prop}.
We then fit the three models to each resampled dataset, and repeat this test 100 times for each model. 
As for the synthetic model calculations, we take our confidence interval from the age range containing 67\% of the best-fit ages, and give the results in Table~\ref{tab:res}. 
This method estimates the uncertainty due to stellar parameter estimation. 
For both Zero Torque fits and the log fit to the Standard model, resampling within the mass uncertainties does not produce significantly different results: $>67\%$ of these tests produce the same best-fit age as our initial analysis. 
In the other three cases, the confidence intervals from mass resampling are smaller than those found by drawing synthetic observations from the model. 

\begin{figure}[t!]
\centerline{\includegraphics[width=\columnwidth]{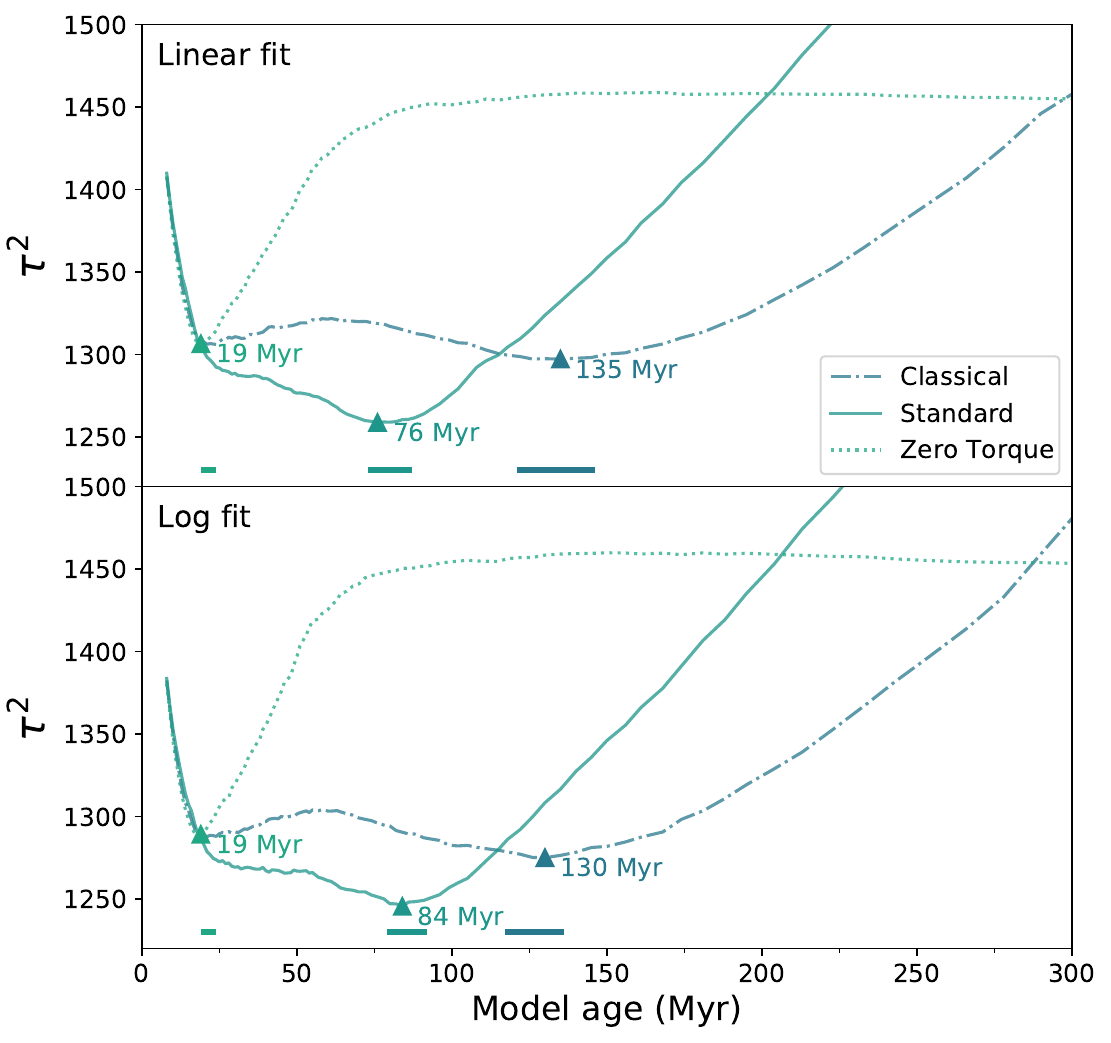}}
\caption{\tausq vs age for all three of our models, from both the linear fit (top panel) and log fit (bottom panel). 
Solid lines show results from the standard model, dot-dashed lines from the classical model, and dotted lines from the zero-torque model. 
The best-fit age from each model is indicated by a triangle point, and the confidence interval is shown by the corresponding solid line at the bottom of the figure. 
See Section~\ref{gof} for details on calculations, including for the confidence intervals. 
The standard-torque model gives the best fit (i.e., smallest \tausq\ value), and the resulting age is closer to the expected 25-55~Myr age than the classical-torque model. 
The results from the three models are statistically inconsistent with each other, indicating that using the \tausq\ method with this ZAMS sample can effectively discriminate between different wind parameterizations.
}
\label{fig:tausq_tracks}
\end{figure}

Statistical uncertainties from the model and the fitting procedure itself therefore dominate our results. 
The resulting confidence intervals are shown at the bottom of each panel in Figure~\ref{fig:tausq_tracks}, and listed in Table~\ref{tab:res}.
The difference in ages between the linear and log fits could be due to the collapsing of detail at \prot$<1$~d in the linear fit. 
Note that the confidence intervals for both fits to the synthetic dataset drawn from the Zero Torque model do not include the best-fit age, even though the confidence interval is based on the model at the best-fit age. 
The confidence intervals are consistent between the log and linear fits for each torque model.

\begin{deluxetable}{lccc}
    \tabletypesize{\footnotesize}
    \tablewidth{0pt}
    \tablecaption{Results of \tausq\ fits to ZAMS \prot \label{tab:res}}
    \tablehead{
        \colhead{} & \colhead{Best-fit age} & \multicolumn{2}{c}{Confidence Intervals (Myr)}\\        \colhead{Model} & \colhead{(Myr)} & \colhead{Model}& \colhead{Mass}
    } 
    \startdata
    \cutinhead{Linear fit}
    Zero Torque & 19 & 20-23\tablenotemark{a} & 19 \\
    Classical & 135 & 122-145 & 126-135 \\
    Standard & 76 & 74-86  & 76-80 \\
    \cutinhead{Log fit}
    Zero Torque & 19 & 20-23\tablenotemark{a} & 19  \\
    Classical & 130 & 118-135 & 126-130 \\
    Standard & 84 & 80-91 &  84 \\
    \enddata
    \tablecomments{The right two columns give the confidence intervals derived via two different methods, described in Section~\ref{unc}. The ``Model'' confidence interval comes from randomly sampling the best-fit model, then fitting models to that synthetic dataset. The ``Mass'' confidence interval comes from randomly re-drawing the stellar masses from within their uncertainties, and then re-fitting the models.}
    \tablenotetext{a}{Confidence intervals do not include the best-fit model.}
\end{deluxetable}

In contrast, the results from each torque model are statistically inconsistent with the results from the other two models. 
This inconsistency between models holds true even if we consider every result from the synthetic datasets (not just the best 67\%), indicating that the \tausq\ method combined with our dataset can clearly distinguish between different models for angular momentum evolution.

\subsection{Discussion}

The \tausq\ model-data comparison gives several insights into the evolution of near-solar-mass stars as they approach the ZAMS.
First, the zero-torque model has the worst fit to the data, which indicates that the data are inconsistent with the prediction of both angular momentum conservation and solid body rotation.
A significant amount of angular momentum must therefore be lost during the pre-main-sequence spin-up phase, and/or a significant amount of angular momentum becomes hidden in the stellar interior by \agerange; redistribution of angular momentum is not included in our models. 

We have only tested two different models here, and future work should explore a wider range of models assumptions and parameters.
Of note, even the best fitting model yields an age ($\approx80$~Myr) significantly older than these clusters likely true ages of \agerange.
Thus, even the best fitting model does not appear to be a ``good'' fit to the new data. 

The model fit may be improved by exploring alternate formulations for the stellar-wind torque and/or by including and exploring the effects of internal angular momentum redistribution.
Approaches that simultaneously model the interior angular momentum transport and external torques, such as \citet{amard2019} and \citet{gossage2021}, may be needed. Unfortunately, however, these models are more complex and have not yet been developed to produce the 2D probability distribution necessary for \tausq\ fitting.

In addition, the age of the best fit models will also be degenerate with the adopted age of the model initial conditions.
We follow \citet{rebull2018} in adopting 8~Myr for the age of Upper Sco, but these authors also note literature values between 3-10~Myr in their introduction. 
If the true age of Upper Sco is younger or older than the adopted age of 8~Myr, then the models would require more or less angular momentum evolution by the ZAMS age.
Future modeling work, and especially comparisons between models, should more thoroughly consider the impact of initial conditions on comparisons with actual data.

Finally, we note that we have made no attempt to remove potential binary systems from our sample. While the renormalized unit weight error (RUWE) from Gaia can identify binaries, it only does so within a fairly narrow range of separations ($0.1\arcsec\lapprox\rho\lapprox1\arcsec$) and magnitudes ($\Delta G\lapprox3$) \citep[e.g.,][]{wood2021}. At the distance of our target clusters, this is sensitive to the peak of the expected period distribution for G stars \citep{duchene2013}, but leaves out both closer and wider systems that have the potential for tidal and disk interactions \citep[e.g.,][]{zahn2008,cieza2009}. Many rapid rotators in our sample have RUWE$\approx1$, supporting the likelihood that some binaries are not captured by this statistic. In addition, the initial conditions for our rotation models are taken from the Upper Sco survey by \citet{rebull2018}, which has not been filtered for binaries either. For now, therefore, we believe our choice not to filter for binaries is reasonable. However, there is strong motivation for binary surveys in these clusters, particularly spectroscopic surveys for tidally interacting systems.

Despite these caveats, the \tausq\ method reveals a very clear difference between the two torque models. 
Both the smaller \tausq\ value and the closer-to-realistic age indicate that the standard model is a much better fit than the classical one.
Our new dataset can clearly discriminate between different models, and will be a valuable benchmark for modeling the early evolution of near-solar-mass stars.

\section{Conclusions}\label{concl}
We analyze TESS light curves for \totaltargets\ members of five open clusters at the ZAMS, and measure \tessperiods\ rotation periods. 
The clusters considered are IC~2391, IC~2602, NGC~2451A, NGC~2547, and Collinder~135. 
Prior work has focused on late K and M stars, leaving Sun-like stars largely unconstrained at this critical evolutionary point. 
We present \tessperiods\ rotation periods, of which \newsolar\ are new, high-quality detections for Sun-like cluster members. 
This represents an \duplit-fold increase in rotation periods for ZAMS solar-mass stars, compared to the literature. 

We then compare our new dataset to three models for stellar angular momentum evolution using the \tausq\ method \citep{naylor2006,breimann2021}. 
Each model incorporates a different stellar wind torque: zero torque, the model from \citet{matt2015}, and an updated torque that better matches observations of rapid rotators in the Pleiades and Praesepe \citep{breimann2021}.
We show that our data and this statistically robust fitting method can clearly discriminate between different stellar wind parameterizations. 
We also find that magnetic braking and/or internal angular momentum transport have significant impacts on angular momentum evolution even before stars reach the ZAMS. 
Our TESS periods therefore provide a new, invaluable resource for understanding the rotational evolution of Sun-like stars and for constraining the behavior of stellar models at this critical age.

\acknowledgments

S.T.D.~ acknowledges support from the National Aeronautics and Space Administration under Grant No. 80NSSC19K0377 issued through the TESS Guest Investigator program.
S.T.D.~also acknowledges support provided by the NSF through grant AST-1701468. Any opinions, findings, and conclusions or recommendations expressed in this material are those of the authors and do not necessarily reflect the views of the NSF. 
S.P.M.~ and A.A.B.~ acknowledge funding from the European Research Council (ERC), under the European Union’s Horizon 2020 research and innovation program (grant agreement No. 682393 AWESoMeStars).
S.P.M.~ also acknowledges support as a visiting scholar from the Center for Computational Astrophysics at the Flatiron Institute, which is supported by the Simons Foundation.
We acknowledge the support of the Center for Astrophysics $\vert$ Harvard \& Smithsonian (Solar, Stellar, and Planetary Division) and the Banneker Institute during the first year of this analysis. 
We thank R. Jackson, R. Jeffries, S. Meibom, and S. Randich  for comments and advice on our intial \tess\ proposal. 
We also thank D.~Rodriguez and J.~Lewis for assistance with creating TESS sky plots, and L.~Rebull for providing archival data. 
We also thank the anonymous referee for comments which improved the manuscript.

This research has made use of NASA's Astrophysics Data System Bibliographic Services, the SIMBAD database \citep{simbad}, operated at CDS, Strasbourg, France, and the VizieR database of astronomical catalogs \citep{Ochsenbein2000}.

This paper includes data collected by the TESS mission. Funding for the TESS mission is provided by the NASA's Science Mission Directorate.

Some of the data presented in this paper were obtained from the Mikulski Archive for Space Telescopes (MAST). STScI is operated by the Association of Universities for Research in Astronomy, Inc., under NASA contract NAS5-26555. Support for MAST for non-{\it HST} data is provided by the NASA Office of Space Science via grant NNX09AF08G and by other grants and contracts. 
Specifically, we use the TESS High-Level Science Products \dataset[QLP]{https://doi.org/10.17909/t9-r086-e880} and \dataset[CDIPS]{https://doi.org/10.17909/t9-ayd0-k727}.

This work has made use of data from the European Space Agency (ESA) mission
{\it Gaia} (\url{https://www.cosmos.esa.int/gaia}), processed by the {\it Gaia}
Data Processing and Analysis Consortium (DPAC,
\url{https://www.cosmos.esa.int/web/gaia/dpac/consortium}). Funding for the DPAC
has been provided by national institutions, in particular the institutions
participating in the {\it Gaia} Multilateral Agreement.

\vspace{5mm}
\facilities{TESS, Gaia}

\software{Astropy \citep{astropy,astropy2018,astropy2022},
        Astroquery \citep{ginsburg2013},
        AstroML \citep{vanderplas2012,ivezic2013}, 
        Lightkurve,
        Matplotlib
        }

\setlength{\baselineskip}{0.6\baselineskip}
\bibliography{references}
\setlength{\baselineskip}{1.667\baselineskip}

\end{document}

%% file: tab_all_lit_periods.tex
\begin{deluxetable*}{lllDlll}
\tabletypesize{\footnotesize}
\tablewidth{0pt}
 \tablecaption{Literature periods for cluster stars, crossmatched to current identifiers
 \label{tab:lit_periods}}
 \tablehead{
\colhead{Name} & \colhead{Cluster} & \colhead{Source} & \twocolhead{LitPeriod (d)} & \colhead{TIC} & \colhead{GaiaDR2} & \colhead{SimbadName}} 
 \decimals
\startdata
VXR 12 & IC 2391 & patten1996 & 3.86 & 93549309 & 5318545521198976000 & VXR PSPC 12 \\
VXR 14 & IC 2391 & patten1996 & 1.32 & 93551206 & 5318096125872352768 & VXR PSPC 14 \\
VXR 35a & IC 2391 & patten1996 & 0.527 & 93833881 & 5318474941990522368 & VXR PSPC 35a \\
VXR 38a & IC 2391 & patten1996 & 2.78 & 93832681 & 5318501334573605504 & VXR PSPC 38a \\
VXR 41 & IC 2391 & patten1996 & 5.8 & 93832296 & 5318504426950057728 & VXR PSPC 41 \\
VXR 42a & IC 2391 & patten1996 & 1.81 & 93912428 & 5318503533597353856 & VXR PSPC 42a \\
VXR 45a & IC 2391 & patten1996 & 0.223 & 93912319 & 5318500303781947776 & VXR PSPC 45a \\
VXR 47 & IC 2391 & patten1996 & 0.258 & 93911997 & 5318498688873909632 & VXR PSPC 47 \\
VXR 60a & IC 2391 & patten1996 & 0.93 & 812594503 & 5318510336826176384 & Cl* IC 2391 SHJM 4 \\
VXR 60b & IC 2391 & patten1996 & 0.212 & 94184691 & 5318510332522027008 & Cl* IC 2391 SHJM 5 \\
\enddata
\tablecomments{This table is available in its entirety in a machine-readable form in the online journal. A portion is shown here for guidance regarding its form and content. }
\tablerefs{patten1996,patten1996-1,barnes1999,irwin2008,messina2011}

\end{deluxetable*}
